\documentclass[%
 reprint,
superscriptaddress,
 amsmath,amssymb,
 aps,
]{revtex4-1}

\usepackage{graphicx}
\usepackage{epstopdf}
\usepackage{dcolumn}
\usepackage{bm}
\usepackage{color}
\usepackage[caption=false]{subfig}
\usepackage{xspace}
\usepackage{epsfig}
\usepackage{multirow}
\usepackage[]{units}
\usepackage{cleveref}
\usepackage{hyperref}
\usepackage[mathlines]{lineno}
\input symbols

\begin{document}

\preprint{XXX}

\title{Measurement of the intrinsic electron neutrino component in the
  T2K neutrino beam with the ND280 detector}


\newcommand{\INSTC}{\affiliation{University of Alberta, Centre for Particle Physics, Department of Physics, Edmonton, Alberta, Canada}}
\newcommand{\INSTEE}{\affiliation{University of Bern, Albert Einstein Center for Fundamental Physics, Laboratory for High Energy Physics (LHEP), Bern, Switzerland}}
\newcommand{\INSTFE}{\affiliation{Boston University, Department of Physics, Boston, Massachusetts, U.S.A.}}
\newcommand{\INSTD}{\affiliation{University of British Columbia, Department of Physics and Astronomy, Vancouver, British Columbia, Canada}}
\newcommand{\INSTGA}{\affiliation{University of California, Irvine, Department of Physics and Astronomy, Irvine, California, U.S.A.}}
\newcommand{\INSTI}{\affiliation{IRFU, CEA Saclay, Gif-sur-Yvette, France}}
\newcommand{\INSTGB}{\affiliation{University of Colorado at Boulder, Department of Physics, Boulder, Colorado, U.S.A.}}
\newcommand{\INSTFG}{\affiliation{Colorado State University, Department of Physics, Fort Collins, Colorado, U.S.A.}}
\newcommand{\INSTFH}{\affiliation{Duke University, Department of Physics, Durham, North Carolina, U.S.A.}}
\newcommand{\INSTBA}{\affiliation{Ecole Polytechnique, IN2P3-CNRS, Laboratoire Leprince-Ringuet, Palaiseau, France }}
\newcommand{\INSTEF}{\affiliation{ETH Zurich, Institute for Particle Physics, Zurich, Switzerland}}
\newcommand{\INSTEG}{\affiliation{University of Geneva, Section de Physique, DPNC, Geneva, Switzerland}}
\newcommand{\INSTDG}{\affiliation{H. Niewodniczanski Institute of Nuclear Physics PAN, Cracow, Poland}}
\newcommand{\INSTCB}{\affiliation{High Energy Accelerator Research Organization (KEK), Tsukuba, Ibaraki, Japan}}
\newcommand{\INSTED}{\affiliation{Institut de Fisica d'Altes Energies (IFAE), Bellaterra (Barcelona), Spain}}
\newcommand{\INSTEC}{\affiliation{IFIC (CSIC \& University of Valencia), Valencia, Spain}}
\newcommand{\INSTEI}{\affiliation{Imperial College London, Department of Physics, London, United Kingdom}}
\newcommand{\INSTGF}{\affiliation{INFN Sezione di Bari and Universit\`a e Politecnico di Bari, Dipartimento Interuniversitario di Fisica, Bari, Italy}}
\newcommand{\INSTBE}{\affiliation{INFN Sezione di Napoli and Universit\`a di Napoli, Dipartimento di Fisica, Napoli, Italy}}
\newcommand{\INSTBF}{\affiliation{INFN Sezione di Padova and Universit\`a di Padova, Dipartimento di Fisica, Padova, Italy}}
\newcommand{\INSTBD}{\affiliation{INFN Sezione di Roma and Universit\`a di Roma ``La Sapienza'', Roma, Italy}}
\newcommand{\INSTEB}{\affiliation{Institute for Nuclear Research of the Russian Academy of Sciences, Moscow, Russia}}
\newcommand{\INSTHA}{\affiliation{Kavli Institute for the Physics and Mathematics of the Universe (WPI), Todai Institutes for Advanced Study, University of Tokyo, Kashiwa, Chiba, Japan}}
\newcommand{\INSTCC}{\affiliation{Kobe University, Kobe, Japan}}
\newcommand{\INSTCD}{\affiliation{Kyoto University, Department of Physics, Kyoto, Japan}}
\newcommand{\INSTEJ}{\affiliation{Lancaster University, Physics Department, Lancaster, United Kingdom}}
\newcommand{\INSTFC}{\affiliation{University of Liverpool, Department of Physics, Liverpool, United Kingdom}}
\newcommand{\INSTFI}{\affiliation{Louisiana State University, Department of Physics and Astronomy, Baton Rouge, Louisiana, U.S.A.}}
\newcommand{\INSTJ}{\affiliation{Universit\'e de Lyon, Universit\'e Claude Bernard Lyon 1, IPN Lyon (IN2P3), Villeurbanne, France}}
\newcommand{\INSTCE}{\affiliation{Miyagi University of Education, Department of Physics, Sendai, Japan}}
\newcommand{\INSTDF}{\affiliation{National Centre for Nuclear Research, Warsaw, Poland}}
\newcommand{\INSTFJ}{\affiliation{State University of New York at Stony Brook, Department of Physics and Astronomy, Stony Brook, New York, U.S.A.}}
\newcommand{\INSTGJ}{\affiliation{Okayama University, Department of Physics, Okayama, Japan}}
\newcommand{\INSTCF}{\affiliation{Osaka City University, Department of Physics, Osaka, Japan}}
\newcommand{\INSTGG}{\affiliation{Oxford University, Department of Physics, Oxford, United Kingdom}}
\newcommand{\INSTBB}{\affiliation{UPMC, Universit\'e Paris Diderot, CNRS/IN2P3, Laboratoire de Physique Nucl\'eaire et de Hautes Energies (LPNHE), Paris, France}}
\newcommand{\INSTGC}{\affiliation{University of Pittsburgh, Department of Physics and Astronomy, Pittsburgh, Pennsylvania, U.S.A.}}
\newcommand{\INSTFA}{\affiliation{Queen Mary University of London, School of Physics and Astronomy, London, United Kingdom}}
\newcommand{\INSTE}{\affiliation{University of Regina, Department of Physics, Regina, Saskatchewan, Canada}}
\newcommand{\INSTGD}{\affiliation{University of Rochester, Department of Physics and Astronomy, Rochester, New York, U.S.A.}}
\newcommand{\INSTBC}{\affiliation{RWTH Aachen University, III. Physikalisches Institut, Aachen, Germany}}
\newcommand{\INSTFB}{\affiliation{University of Sheffield, Department of Physics and Astronomy, Sheffield, United Kingdom}}
\newcommand{\INSTDI}{\affiliation{University of Silesia, Institute of Physics, Katowice, Poland}}
\newcommand{\INSTEH}{\affiliation{STFC, Rutherford Appleton Laboratory, Harwell Oxford,  and  Daresbury Laboratory, Warrington, United Kingdom}}
\newcommand{\INSTCH}{\affiliation{University of Tokyo, Department of Physics, Tokyo, Japan}}
\newcommand{\INSTBJ}{\affiliation{University of Tokyo, Institute for Cosmic Ray Research, Kamioka Observatory, Kamioka, Japan}}
\newcommand{\INSTCG}{\affiliation{University of Tokyo, Institute for Cosmic Ray Research, Research Center for Cosmic Neutrinos, Kashiwa, Japan}}
\newcommand{\INSTGI}{\affiliation{Tokyo Metropolitan University, Department of Physics, Tokyo, Japan}}
\newcommand{\INSTF}{\affiliation{University of Toronto, Department of Physics, Toronto, Ontario, Canada}}
\newcommand{\INSTB}{\affiliation{TRIUMF, Vancouver, British Columbia, Canada}}
\newcommand{\INSTG}{\affiliation{University of Victoria, Department of Physics and Astronomy, Victoria, British Columbia, Canada}}
\newcommand{\INSTDJ}{\affiliation{University of Warsaw, Faculty of Physics, Warsaw, Poland}}
\newcommand{\INSTDH}{\affiliation{Warsaw University of Technology, Institute of Radioelectronics, Warsaw, Poland}}
\newcommand{\INSTFD}{\affiliation{University of Warwick, Department of Physics, Coventry, United Kingdom}}
\newcommand{\INSTGE}{\affiliation{University of Washington, Department of Physics, Seattle, Washington, U.S.A.}}
\newcommand{\INSTGH}{\affiliation{University of Winnipeg, Department of Physics, Winnipeg, Manitoba, Canada}}
\newcommand{\INSTEA}{\affiliation{Wroclaw University, Faculty of Physics and Astronomy, Wroclaw, Poland}}
\newcommand{\INSTH}{\affiliation{York University, Department of Physics and Astronomy, Toronto, Ontario, Canada}}

\INSTC
\INSTEE
\INSTFE
\INSTD
\INSTGA
\INSTI
\INSTGB
\INSTFG
\INSTFH
\INSTBA
\INSTEF
\INSTEG
\INSTDG
\INSTCB
\INSTED
\INSTEC
\INSTEI
\INSTGF
\INSTBE
\INSTBF
\INSTBD
\INSTEB
\INSTHA
\INSTCC
\INSTCD
\INSTEJ
\INSTFC
\INSTFI
\INSTJ
\INSTCE
\INSTDF
\INSTFJ
\INSTGJ
\INSTCF
\INSTGG
\INSTBB
\INSTGC
\INSTFA
\INSTE
\INSTGD
\INSTBC
\INSTFB
\INSTDI
\INSTEH
\INSTCH
\INSTBJ
\INSTCG
\INSTGI
\INSTF
\INSTB
\INSTG
\INSTDJ
\INSTDH
\INSTFD
\INSTGE
\INSTGH
\INSTEA
\INSTH

\author{K.\,Abe}\INSTBJ
\author{J.\,Adam}\INSTFJ
\author{H.\,Aihara}\INSTCH\INSTHA
\author{T.\,Akiri}\INSTFH
\author{C.\,Andreopoulos}\INSTEH
\author{S.\,Aoki}\INSTCC
\author{A.\,Ariga}\INSTEE
\author{T.\,Ariga}\INSTEE
\author{S.\,Assylbekov}\INSTFG
\author{D.\,Autiero}\INSTJ
% ...

\collaboration{The T2K Collaboration}\noaffiliation

\date{\today}

\begin{abstract}
The T2K experiment has reported the first observation of the
appearance of electron neutrinos in a muon
neutrino beam. The
main and irreducible background to the appearance signal comes from the presence in
the neutrino beam of a small intrinsic component of electron neutrinos
originating from muon and kaon decays. In T2K, this component is expected to represent 1.2\% of the
total neutrino flux. A measurement of this component using the near detector (ND280),
located 280~m from the target, is presented. The charged current
interactions of electron neutrinos are selected by combining the particle identification
capabilities of both the time projection chambers and 
electromagnetic calorimeters of ND280. The measured ratio between the
observed electron neutrino beam component and the prediction is $1.01
\pm 0.10$ providing a direct confirmation of the neutrino fluxes and
neutrino cross section modeling used for T2K neutrino oscillation
analyses. Electron neutrinos coming from muons and kaons decay are also separately measured,
resulting in a ratio with respect to the prediction of $0.68 \pm 0.30$
and $1.10 \pm 0.14$, respectively.


\end{abstract}

\pacs{Valid PACS appear here}
\maketitle


\section{\label{sec:intro} Introduction}

The T2K (Tokai-to-Kamioka) experiment~\cite{Abe:2011ks} is a long baseline neutrino
oscillation experiment that uses an intense muon neutrino beam
produced at the Japan Proton Accelerator Research Complex (J-PARC) in
Tokai. The primary goals of the experiment are the precise determination of the
oscillation parameter $\theta_{13}$ via electron neutrino
appearance, and of the parameters $\theta_{23}$ and $\Delta m^2_{32}$
via muon neutrino disappearance. Neutrino interactions are observed at
a near detector, ND280, where the flavour composition of the incoming
neutrino flux is not expected
to be affected by oscillation, and at the far detector,
Super-Kamiokande (SK), where oscillation significantly affects the
composition.

The T2K baseline, the neutrino beam configuration, and the ability of the far detector to distinguish
electrons from muons results in excellent sensitivity for \nue
appearance. The $\num \rightarrow \nue$ oscillation probability depends on
$\theta_{13}$, and on sub-leading effects that depend on the \dcp
phase and on the mass hierarchy~\cite{PhysRevD.56.3093}.
Recently T2K reported the first observation of electron neutrino
appearance with a 7.3~$\sigma$ significance, by observing 28 electron
neutrino events compared to a background expectation of $4.92\pm0.55$ events
for $\theta_{13}=0$~\cite{Abe:2013hdq}. Among those background events, 3.2
are expected to be due to the intrinsic $\nu_e$ beam component, an
irreducible background to the electron neutrino appearance
search. 

The intrinsic \nue in the beam are an unavoidable product of conventional neutrino
beams where pions and kaons, produced by the interaction of a proton
beam with a target, decay to form a \num beam. When the muons and kaons decay, a small component of \nue is produced in addition to the \num. In the T2K case \nue are expected to
represent about 1.2\% of the total neutrino
flux~\cite{PhysRevD.87.012001}. This component 
will be the main source of background for all the proposed long-baseline neutrino oscillation
experiments~\cite{Abe:2011ts, Adams:2013qkq, Stahl:2012exa}
aiming to measure CP violation in the leptonic sector by precisely measuring
\nue (\nueb) appearance in a \num (\numb) beam. A direct measurement of this component performed
at ND280 is presented in this paper.

ND280 is a magnetized detector located at a distance of 280~m from
the target. For this analysis neutrino charged current (CC)
interactions in the Fine Grained Detectors (FGDs~\cite{Amaudruz:2012pe}) are selected. The combination of the
particle identification (PID) capabilities of three Time Projection
Chambers (TPC~\cite{Abgrall:2010hi}) and a set of Electromagnetic Calorimeters (ECals~\cite{Allan:2013ofa}) is
used to distinguish electrons from muons, allowing the selection of a
clean sample of \nue CC interactions with a
purity of about 65\%. The background is dominated by photon
conversions producing \epem pairs in the FGD.

In the T2K oscillation analyses the measurement of the spectra 
of \num CC interactions at ND280 is used to constrain the uncertainties
on the unoscillated neutrino fluxes and on the neutrino cross section
parameters. 
The \num CC sample constrains also the \nue flux and cross section
because of the significant correlations between the \num and the \nue
fluxes, which originate from the same hadrons. The \num
and \nue cross sections are expected to be the same, except for
radiative corrections and the different lepton mass~\cite{Day:2012gb}.

The beam \nue component is directly measured at
ND280 and it is compared with the expectations when fluxes and
cross section uncertainties are constrained by the same fit to the
ND280 \num CC sample used for the oscillation analyses. 
This measurement directly confirms the validity of the procedure used in all T2K oscillation analyses.

The paper is organized as follows. In Section \ref{sec:t2k} the T2K
experiment, the flux prediction and the neutrino interaction
cross section model are described, and in Section \ref{sec:nd280} 
ND280 and the detectors used in this analysis are
introduced. The electron neutrino selection is then described in
Section \ref{sec:nuesel} and the control of the backgrounds entering
the analysis is shown in Section \ref{sec:background}. A description
of the systematics is given in Section \ref{sec:systematics} and
the fit used to extract the beam \nue component is shown in Section
\ref{sec:fit}. Finally the results are given in Section
\ref{sec:results} and a summary in Section
\ref{sec:summary}.

\section{\label{sec:t2k} Experimental overview}

T2K is the first long baseline experiment designed to observe electron
neutrino appearance in a nearly pure muon neutrino beam. The neutrino
beam is produced by the J-PARC accelerator complex where protons are
accelerated up to 31 GeV/c before being extracted in 5~$\mu s$ long
spills with a repetition rate that has been decreased from 3.6~s to
2.6~s over the data-taking periods. The spill consists of 8 bunches (6
during the first data-taking period), each 15~ns wide. The protons
strike a 91.4 cm long graphite target, producing hadrons, mainly pions
and kaons. The positively charged particles are focused by a series of
three magnetic horns operating at 250 kA before entering a 96 m long
decay volume where they decay producing mainly muon neutrinos. A small
fraction of the kaons, and the muons produced by pion decay, can also
decay producing electron neutrinos. Most of the surviving charged
particles are stopped in a beam dump at the end of the decay volume. A
muon monitor (MUMON) situated downstream of the beam dump measures the
profile of high energy muons not stopped by the beam dump, monitoring
the stability of the beam intensity and the direction of the beam. The
neutrinos are sampled 280 m and 295 km from the target, at the ND280
near detector and Super-Kamiokande (SK) far detector, respectively.

The direction of the proton beam and the axis of the target and horns is
$2.5^{\circ}$ away from the direction to SK, giving a narrow band \num
beam peaked at 0.6~\gev towards SK. This corresponds to the
oscillation maximum for the 295 km baseline.
T2K is the
first experiment designed to use this configuration, called the off-axis technique~\cite{offaxis}. This configuration also has the advantage
of reducing the beam \nue component in the oscillation region and the
high energy neutrino flux which contributes to backgrounds in the
oscillation analyses.

The near detector complex is comprised of an on-axis detector
(INGRID~\cite{Otani2010368}) and an off-axis detector (ND280) that will be described in detail in the next section. 
SK is a 50 kt cylindrical water Cherenkov
detector. 
The water tank is optically separated into two concentric detectors, an inner detector
(ID) and an outer detector
(OD), both instrumented with photomultipliers. Charged particles
emitted from neutrino interactions produce photons through the
Cherenkov effect and ring-shaped patterns are detected on the walls by
the photomultipliers. The ring pattern is used to identify the
type of the particle. Until
recently, the main backgrounds to the electron neutrino appearance
searches came from the intrinsic beam $\nu_e$ and from neutral current
(NC) interactions in which a $\pi^0$ in the final state (NC$\pi^0$) is produced
but only one electron-like ring is reconstructed~\cite{Abe:2011sj, Abe:2013xua}. A new reconstruction
algorithm~\cite{Abe:2013hdq} has been developed to substantially reduce the
NC$\pi^0$ background, leaving the intrinsic beam
$\nu_e$ component as the main background to the \nue appearance analysis.

In this paper a direct measurement of the \nue beam component is presented. All the
data collected between January 2010 and May 2013 are used
for the analysis. The data are subdivided into different run periods as shown in
Tab.~\ref{tab:pot}. A small fraction of Run III data ($\sim15\%$) was collected with
magnetic horns operating at 205 kA instead of the nominal 250 kA, while
for Run I data only one subset of the ECal, the downstream module, was
installed and operated. The remaining modules were installed and commissioned before the start of Run II. The
simulated data used in this analysis corresponds to more than ten times the
\pot of the data, and the various experimental
conditions of the different data taking periods are reproduced.

\begin{table}
  \caption{\label{tab:pot} Definition of T2K runs and the number of 
    protons on target (\pot) used in the analysis.}
\begin{ruledtabular}
 \begin{tabular}{ccd}
T2K run & Dates & \multicolumn{1}{c}{ND280 \pot ($\times 10^{19}$)} \\
\hline 
  Run I & Jan. 2010 -- Jun. 2010 &  1.7 \\ 
  Run II & Nov. 2010 -- Mar. 2011 & 7.9\\
  Run III & Mar. 2012 -- Jun. 2012 & 15.6 \\
  Run IV & Jan. 2013 -- May. 2013 & 33.8 \\ 
\hline
Total & Jan. 2010 -- May 2013 & 59.0 \\
 \end{tabular}
\end{ruledtabular}
\end{table}

\subsection{\label{sec:flux} Flux prediction}

A good knowledge of the initial neutrino fluxes at ND280 and at SK is
fundamental for all the physics analyses in T2K. The \num (\nue)
components of the beam are mainly produced through charged pion (muon)
and kaon decays.

In the T2K simulation the interaction of the primary proton beam and
the propagation of secondary particles in the carbon target are
simulated with FLUKA~\cite{Ferrari:2005zk}. The flux prediction
is based on the hadron production measurements performed by
NA61/SHINE, a fixed target experiment at the CERN SPS in which a
proton beam of the same energy as the T2K beam interacts with a thin
carbon target (2 cm long) or with a T2K replica
target (91.4 cm long)~\cite{Abgrall:2014xwa}. The charged hadrons produced in the proton-carbon interactions are tracked by a system of TPCs, and their
production cross sections as functions of momentum and angle are
measured. The NA61/SHINE measurements cover most of the relevant
hadron production phase space for the T2K flux. The observed production cross section of pions~\cite{Abgrall:2011ae}
and kaons~\cite{Abgrall:2011ts} on the thin target are used in
the T2K flux simulation to reduce the uncertainties on the
flux prediction~\cite{PhysRevD.87.012001}.
Measurements from other experiments (Eichten et
al.~\cite{eichten} and Allaby et al.~\cite{allaby}) are used to
reduce the uncertainty of the particle production in the
region not covered by NA61/SHINE.

\begin{figure}
\includegraphics[width=\linewidth]{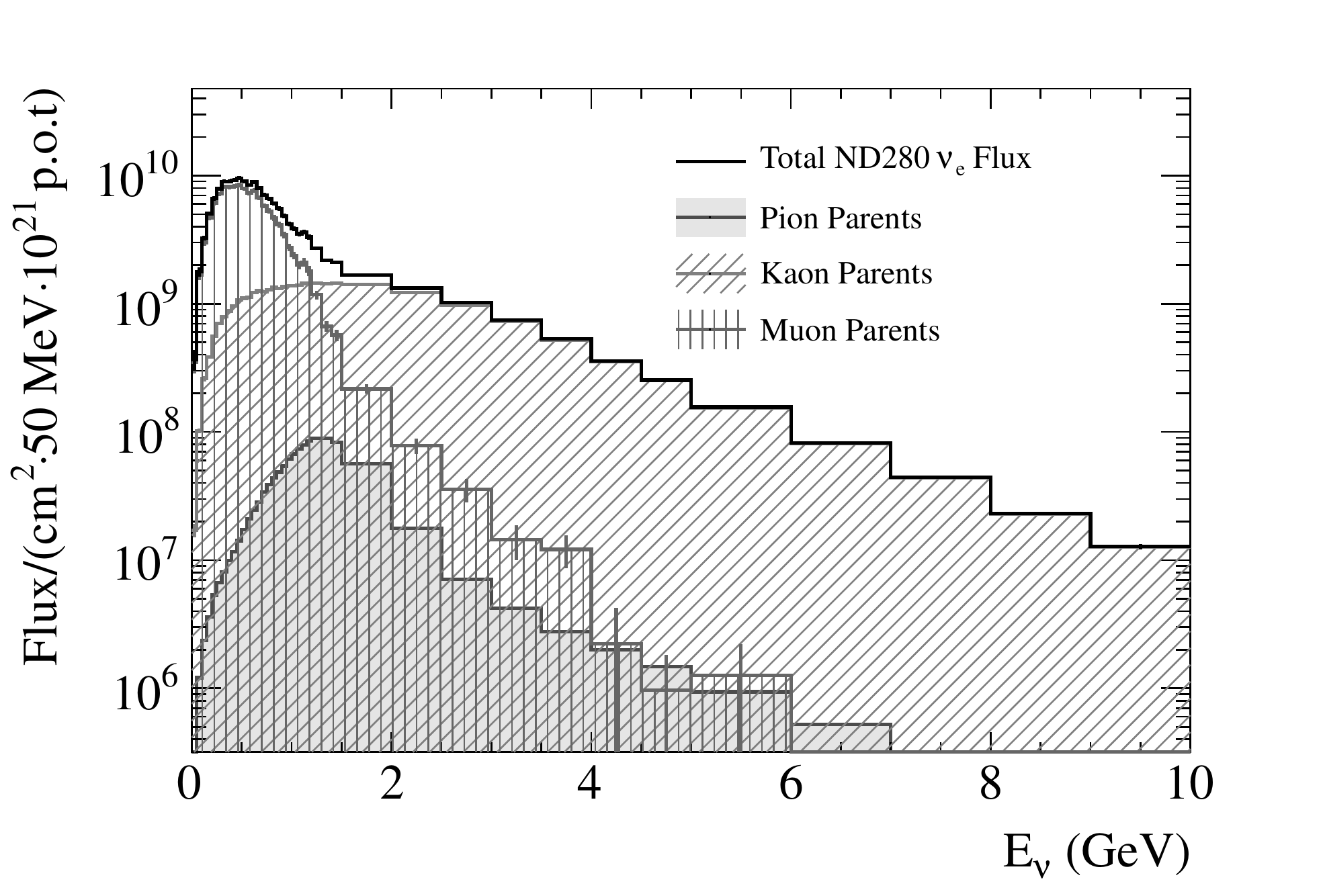}
\caption{\label{fig:t2kflux} The prediction of the \nue flux at ND280
  broken down by the neutrino parent particle type~\cite{PhysRevD.87.012001}.}
\end{figure}

\begin{figure}
\includegraphics[width=\linewidth]{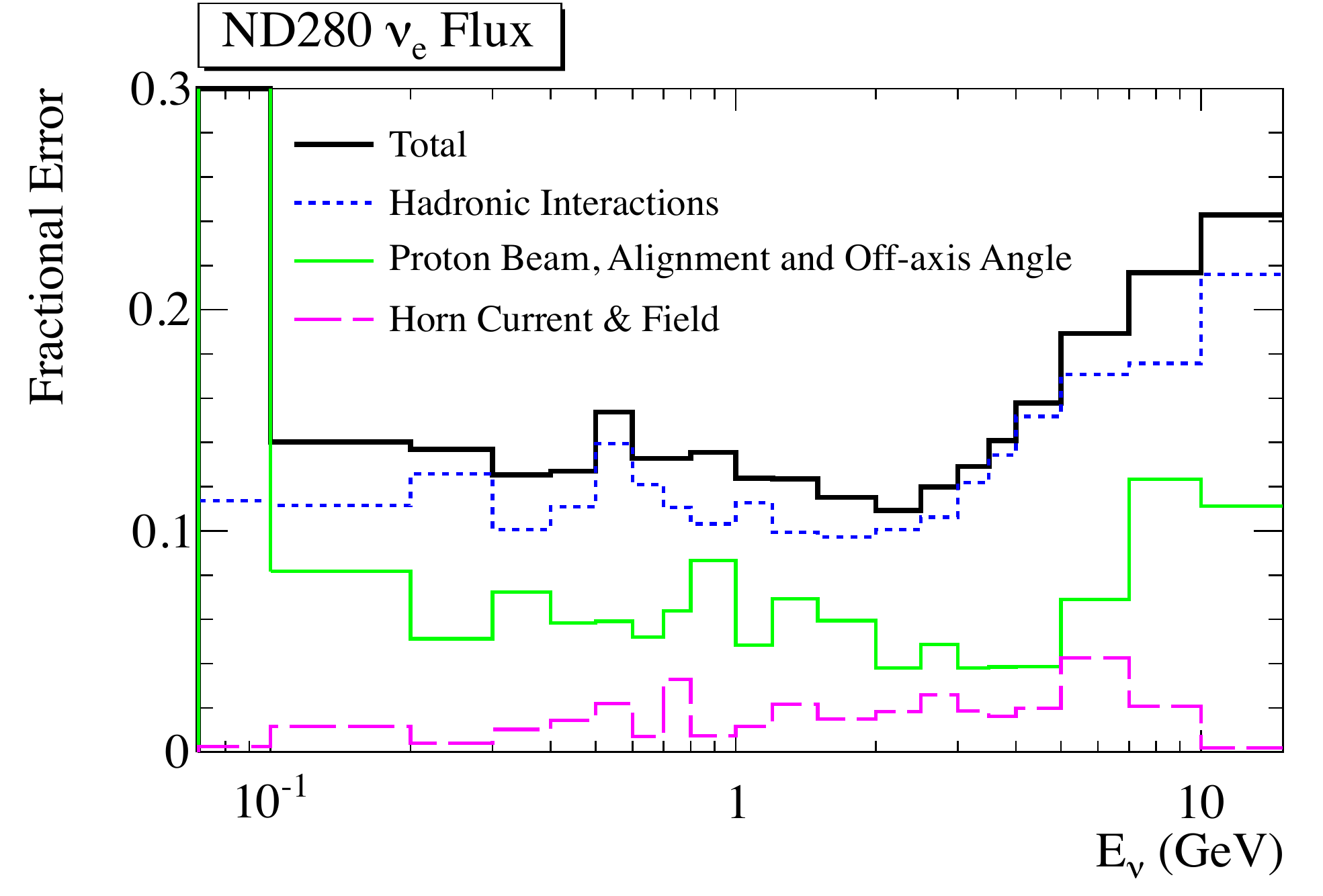}
\caption{\label{fig:t2ksyst} The \nue flux uncertainties at ND280~\cite{PhysRevD.87.012001}.}
\end{figure}

The propagation of particles through the elements of the beamline is simulated with
GEANT3~\cite{GEANT3}. The particles are propagated through the horns'
magnetic field and may interact with the surrounding materials. Particle decays into
neutrinos are simulated as well as the interactions in the decay volume and the
beam dump. The modeling of hadronic interactions is done
using the GCALOR model~\cite{GCALOR}. The beam
direction, its intensity and the beam profile are measured by
the INGRID and MUMON detectors. The neutrino fluxes are described by a
covariance matrix in bins of neutrino energy and type. 
The uncertainty on the \num flux is below 12\% for neutrino
energies around 0.6~\gev. The expected
\nue flux and its uncertainty at ND280 are shown in
Figs.~\ref{fig:t2kflux} and~\ref{fig:t2ksyst}. 
Most of the intrinsic beam \nue component comes
from the decay of $\mu^+$ ($\mu^+ \rightarrow e^{+} \bar{\nu}_{\mu}
\nu_e$) produced by the pions' decay and from charged and neutral
kaons. The charged kaons produce \nue via the decays $K^+ \rightarrow \pi^0
e^+ \nue$ that has a branching ratio of 5.1\%, while the neutral kaons
produce \nue through the decay $K^0_L \rightarrow
\pi^- e^+ \nue$ that has a branching ratio of 40.5\%. The \nue from
muon decays contribute to most of the flux in the low energy region,
below 1.5~\gev, while above that energy almost all of the
\nue flux comes from kaon decays. The \nue from pion decays only
contribute to about 1\% of the total \nue flux (Fig.~\ref{fig:t2kflux}). A more
detailed discussion of the uncertainties contributing to the \nue flux
(Fig.~\ref{fig:t2ksyst}) can be found in~\cite{PhysRevD.87.012001}.
As the physics processes leading to \nue from muon decay and from kaon
decay are different, the analysis presented in this
paper extracts a measurement of their separate contributions to the
flux, as well as the inclusive flux of \nue.

\subsection{\label{sec:xsec} Neutrino interaction model}
 
Neutrino interactions in ND280 are simulated using the
\neut~\cite{NEUT} event generator. This generator covers the range of
neutrino energies from several tens of MeV to hundreds of TeV, and it
simulates the full range of nuclear targets used in ND280. In the simulation, neutrino interactions are generated in the
entire ND280 volume on both active and inactive targets, providing
the necessary information for the signal and for the backgrounds
coming from interactions occurring outside of the ND280 inner detectors. A
complete description of the models used in the T2K
simulation is given in~\cite{CCinc}.

The dominant cross section process at the peak of the T2K beam energy
is Charged Current Quasi-Elastic scattering (CCQE): $\nu_l~+N~\rightarrow
l~+N$, while at higher energies, above the pion production threshold,
single pion production (CC1$\pi$) and deep inelastic scattering (DIS)
contribute to the total charged current cross section.

In \neut, CCQE interactions are simulated using
the model of Llewellyn Smith~\cite{LlewellynSmith:1972}, with the
nuclear effects described by the relativistic Fermi gas model of Smith
and Moniz~\cite{SmithMoniz:1972, SmithMonizErratum}. The form factors
describing the vector and the axial masses are parameterized with
$M_V~=~0.84$ GeV and $M_A~=~1.21$ GeV. The Fermi momentum is set to 217 (225) MeV/c and
the binding energy to 25 (27) MeV for carbon (oxygen).

The pion production is simulated in \neut using the model of Rein and
Sehgal~\cite{ReinSehgal:1981}. Below neutrino energies of 2~\gev, 18 resonances and their
interference terms are simulated. For 20\% of the $\Delta$ resonances
\neut simulates pion-less decay in which the $\Delta$ de-excites
without emitting pions.

Multi-pion and DIS processes are simulated using the GRV98
parton distribution functions~\cite{Gluck:1998xa}. If the invariant mass of
the hadronic system ($W$) is in the range $1.3<W<2.0~\mathrm{GeV/c^2}$ 
only pion multiplicities greater than one are considered to avoid double
counting with the Rein and Sehgal model. For $W>2.0$~GeV/c$^2$
PYTHIA/JETSET~\cite{Sjostrand:1993yb} is used, applying the corrections in the
small $Q^2$ region developed by Bodek and Yang~\cite{Bodek:2003wd}. Additional details on the \neut simulation can be
found in~\cite{Abe:2011ks}.

\subsubsection{\label{sec:xsecsyst} Neutrino interaction uncertainties}

The modeling of the neutrino interactions constitutes an important
source of systematic uncertainties for all T2K analyses. A detailed
description of the uncertainties can be found
in~\cite{Abe:2013xua}. Only a brief summary of the systematic
uncertainties is given here.

\paragraph {CCQE model uncertainty}

Recent measurements of CCQE scattering in the 1 GeV
region~\cite{mb-ccqe} show large discrepancies on the measurement of
the axial mass $M_A^{QE}$ with respect to older measurements. The
strategy that is chosen in T2K analyses is to allow the ND280
$\nu_{\mu}$ CC samples to constrain this parameter, including a large
prior uncertainty ($\sigma_{M_A^{QE}}~=~0.43~\gev$) to account for the
difference between the \neut nominal value and the \neut best-fit to
the \mb data.  Additional degrees of freedom are allowed by three
independent CCQE normalization factors ($x^{QE}_{1,2,3}$) for
different neutrino energy ranges. Below 1.5 GeV an uncertainty of 11\%
is assigned to $x^{QE}_{1}$, corresponding to the uncertainty of the
MiniBooNE flux. The other two normalization factors, $x^{QE}_{2}$ for
$1.5<E_{\nu}<3.5$ GeV and $x^{QE}_{3}$ for $E_{\nu}>3.5$ GeV are given
a prior uncertainty of 30\% to account for the discrepancy between
MiniBooNE and NOMAD data~\cite{Lyubushkin:2008pe}.

\paragraph{Pion production}

For single pion production a joint fit to the MiniBooNE measurements
of charged current single $\pi^+$ production
(CC1$\pi^+$)~\cite{mb-cc1pip}, charged current single $\pi^0$
production (CC1$\pi^0$)~\cite{mb-cc1pi0}, and neutral current single
$\pi^0$ production (NC1$\pi^0$)~\cite{mb-nc1pi0} using \neut has been
performed, varying several parameters.

The parameters varied include the axial mass in the Rein and Sehgal
model $M_A^{RES}$, the normalization of CC1$\pi$
($x^{\mathrm{CC1}\pi}_{1}$ for $E_{\nu}<2.5$~\gev and
$x^{\mathrm{CC1}\pi}_{2}$ for $E_{\nu}>2.5$~\gev), and the
normalization of NC1$\pi^0$ ($x^{\mathrm{NC1}\pi^0}$).

Contributions to the \mb samples from CC multi-pion, NC coherent
interactions, NC charged pion interactions and NC multi-pion are
relatively small and they are included in the analysis described here
with a large prior uncertainty. For charged current coherent pion
production a 100\% normalization uncertainty($x^{\mathrm{CC~coh.}}$),
motivated by the non-observation of the process in the few-GeV energy
range by K2K~\cite{Hasegawa:2005} and SciBooNE~\cite{Hiraide:2008}, is
assigned.  For neutral current charged pion production and all other
NC interactions, including deep inelastic scattering (DIS), a 30\%
normalization uncertainty is introduced ($x^{\mathrm{NC~oth.}}$).

A 20\% uncertainty on the fraction of $\Delta$ that de-excites without
emitting pions ($x^{\pi\mathrm{-less}}$) is also included.


Finally for CC multi-pion/DIS interactions an energy dependent
uncertainty is added ($x^{\mathrm{CC~other}}$), applying a weight $w$
with the form
$w=1+x^{\mathrm{CC~other}}/E_{\nu}(\gev)$. $x^{\mathrm{CC~other}}$ is
allowed to vary around a nominal value of 0 with a prior uncertainty of
0.4.

\paragraph {Nuclear model uncertainties}
\neut models nuclei with a relativistic Fermi gas model (RFG) using
the Fermi momentum $p_F$ determined from electron scattering data. The
uncertainty on $p_F$ is 30 MeV/c, covering possible discrepancies in
the CCQE cross section at low $Q^2$. The uncertainty is applied
independently for interactions on carbon and oxygen targets.

Alternatives to the RFG model of the nuclei are considered by making
comparisons to a spectral function nuclear model implemented in the
NuWro neutrino interaction generator~\cite{Golan:2012rfa}. The
discrepancy in CCQE interaction models using the RFG and spectral
function are assigned as the uncertainty and represented by the
parameter $x_{SF}$, which linearly varies the predicted lepton
kinematics between the RFG ($x_{SF}~=~0$) and spectral function
($x_{SF}~=~1$) models.

\paragraph {Final state interactions (FSI) model tuning}
The \neut FSI model includes parameters which alter the microscopic
pion interaction probabilities in the nucleus. The central value of
these parameters and their uncertainties are determined from fits to
pion scattering data~\cite{dePerio:2011zz}. 

The cross section model parameters and their uncertainties are
summarized in Tab.~\ref{tab:xsec_param_unc}. These uncertainties are
used as prior uncertainties in the fit, along with the flux
uncertainties to the ND280 fit to the $\nu_{\mu}$ CC samples. As will
be shown in Sect.~\ref{sec:physsyst}, the flux systematic
uncertainties are reduced by the measurements of \num CC interactions
in ND280.

\begin{table}
  \caption{\label{tab:xsec_param_unc} The parameters in the \neut
    cross section model along with their nominal values and
    uncertainties prior to the analysis of ND280 data.}
\begin{ruledtabular}
 \begin{tabular}{cdd}
Parameter          & \multicolumn{1}{c}{Nominal Value} & \multicolumn{1}{c}{Uncertainty} \\ 
\hline
$M_A^{QE}$ (GeV)   & 1.21          & 0.45         \\ 
$x^{QE}_{1}$       & 1.00          & 0.11         \\ 
$x^{QE}_{2}$       & 1.00          & 0.30         \\ 
$x^{QE}_{3}$       & 1.00          & 0.30         \\ 
$x_{SF}$           & 0.0           & 1.0          \\
$p_F(^{12}C)$ (MeV/c)   & \multicolumn{1}{c}{217}      & \multicolumn{1}{c}{30}           \\ 
$p_F(^{16}O)$ (MeV/c)   & \multicolumn{1}{c}{225}     & \multicolumn{1}{c}{30}           \\ 
$E_B(^{12}C)$ (MeV)   & \multicolumn{1}{c}{25}      & \multicolumn{1}{c}{3}          \\ 
$E_B(^{16}O)$ (MeV)   & \multicolumn{1}{c}{27}      & \multicolumn{1}{c}{3}           \\ 
$M_A^{RES}$ (GeV)  & 1.41          & 0.11         \\ 
$x^{\mathrm{CC1}\pi}_{1}$   & 1.15          & 0.32         \\ 
$x^{\mathrm{CC1}\pi}_{2}$   & 1.00          & 0.40         \\ 
$x^{\mathrm{NC1}\pi^0}$     & 0.96          & 0.32         \\
$x^{\pi\mathrm{-less}}$     & 0.20           & 0.20          \\ 
$x^{\mathrm{CC~coh.}}$       & 1.00           & 1.00          \\
$x^{\mathrm{NC~other}}$         & 1.00           & 0.30          \\
$x^{\mathrm{CC~other}}$ (GeV)   & 0.00          & 0.40          \\
\end{tabular}
\end{ruledtabular}
\end{table}
 
\section{\label{sec:nd280} The ND280 detector }

The off-axis ND280 detector is a magnetized multi-purpose detector located 
at the same off-axis angle as SK, at a distance of 280 m from the T2K
target. The main purpose of ND280 is to measure the properties of $\nu_{\mu}$
and $\nu_e$ CC interactions before oscillation, reducing
uncertainties in the T2K oscillation analyses. It is also used to
measure neutrino cross sections.

\begin{figure}
\includegraphics[width=\linewidth]{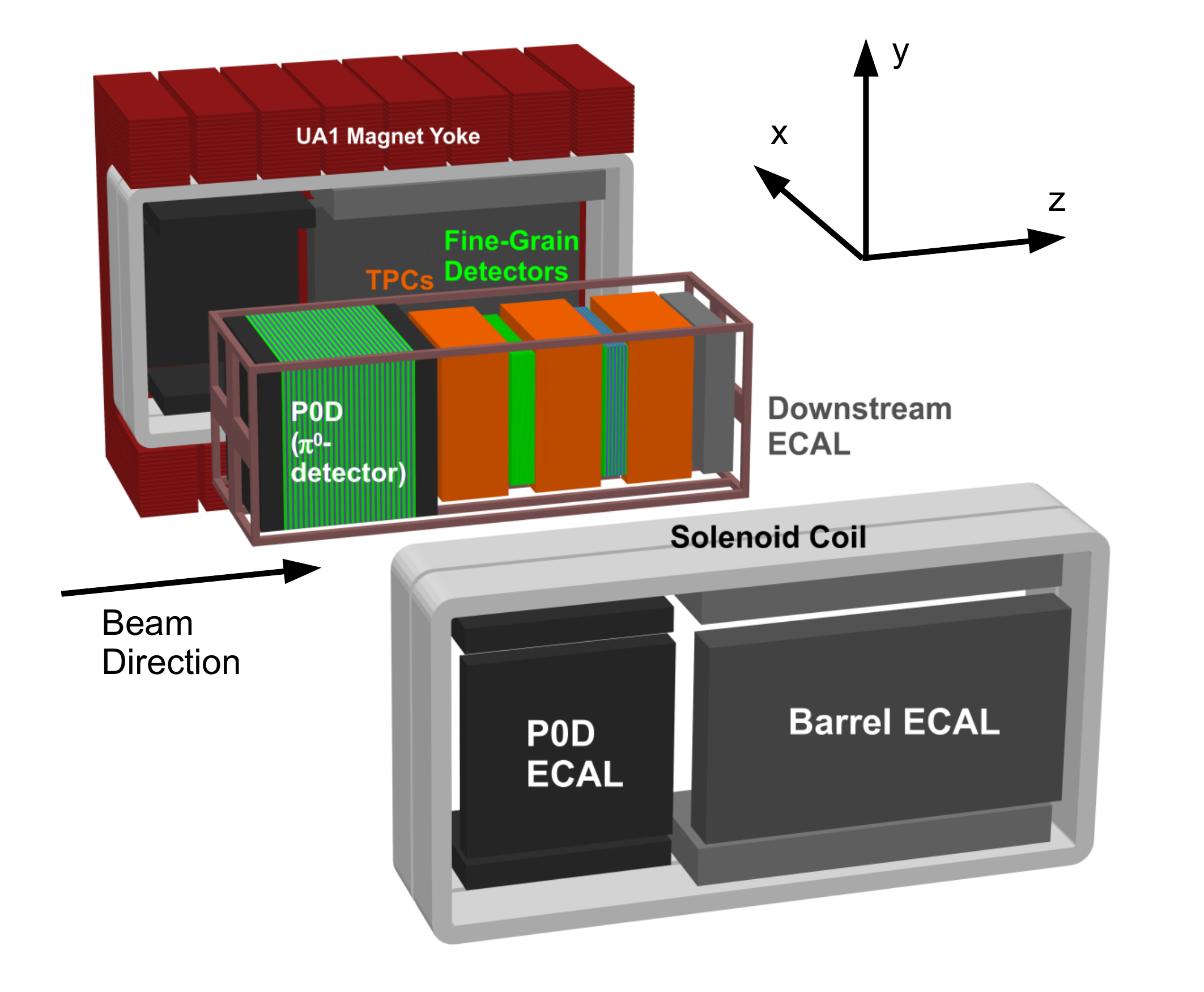}
\caption{\label{fig:nd280layout} A schematic view of the ND280 detector.}
\end{figure}

The layout of ND280 is shown in Fig.~\ref{fig:nd280layout} and a
complete description can be found in~\cite{Abe:2011ks}. It is composed
of a number of sub-detectors installed inside the refurbished
UA1/NOMAD magnet that provides a magnetic field of 0.2 T. The
sub-detectors are side muon range detectors (SMRD~\cite{Aoki:2012mf})
installed in the magnet yokes to track high angle muons, a $\pi^0$
detector (P0D~\cite{Assylbekov201248}) explicitly built to measure
neutrino interactions with the production of $\pi^0$ in the final
state, and a tracking system. The tracking detector is composed of two
fine-grained detectors (FGDs) used as the target for the neutrino
interactions, and three time projection chambers (TPCs). The tracker
and the P0D are surrounded by a set of electromagnetic calorimeters
(ECal). 
In this analysis the tracking detector, Downstream ECal (DsECal) and
the Barrel ECal modules are used. The DsECal is installed downstream
of the tracker system while the Barrel ECal surrounds the tracker and
consists of six different modules (two installed at the top of the
tracker, two at the bottom and one at each side).

The first (upstream) FGD is composed of extruded polystyrene
scintillator bars with layers oriented alternately in the $x$ and $y$
directions (defined in Fig.~\ref{fig:nd280layout}), allowing three
dimensional tracking of the charged particles. The second FGD has the
same structure, but the polystyrene bars are interleaved with water
layers to allow for the measurement of neutrino interactions on water.

The TPCs consist of an inner box filled with Ar:CF$_4$:iC$_4$H$_{10}$
and an outer box filled with CO$_{2}$. Each side of the TPCs is
instrumented with 12 MicroMEGAS modules arranged in two columns. 
Each MicroMEGAS is segmented into 1728 pads arranged in 48 rows and 36
columns, allowing a 3D reconstruction of charged particles produced in
neutrino interactions.

The ECals are sampling calorimeters consisting of layers of 1~cm of
plastic scintillator, divided into bars 4~cm wide, separated by 1~mm
layers of lead. Alternating layers are aligned orthogonally to one
another to provide three dimensional reconstruction of tracks and
showers. The DsECal consists of 34 layers with readout from both ends
of the scintillator bars. The Barrel ECal has 31 layers with readout
from both ends (one end) on the bars parallel (perpendicular) to the
beam direction.

For the analysis described in this paper, neutrino interactions in
both the FGDs are selected by requiring at least one track to enter
the downstream TPC. A combination of TPC and ECal (when available)
particle identification (PID) is used to select electrons, rejecting
most of the muon background produced by the dominant $\nu_{\mu}$ CC
interactions in the FGD.

\subsection{\label{sec:tpcpid} TPC reconstruction and PID performance}

To reconstruct tracks in the TPC, the ionization signals on pads that
exceed a threshold are saved as waveforms.  Waveforms coincident in
time and on adjacent pads in the vertical direction are joined to form
clusters.  Contiguous clusters are then combined to form track
candidates and the kinematic parameters of the track are obtained with
a maximum-likelihood fit to the observed charge distribution. After
track reconstruction in the TPC, signals in the FGD are matched to the
TPC tracks.

The PID in the TPC is based on the measurement of the ionization
produced by charged particles crossing the gas.  To perform particle
identification in the TPC, the ionization in each cluster is corrected
for the track length sampled by the pad column. Using 70\% of the
lowest charge deposits on the pads, a mean ionization value is calculated
and is compared to that expected for particle type $i$ at the
reconstructed momentum.  This comparison is used to form the `pull'
$\delta_i$ (the difference between the measured mean ionization and
the expected one divided by the resolution). The resolution depends on the
number of samples and path length with a typical resolution for muons
of 8\%.

\begin{figure}
\includegraphics[width=\linewidth]{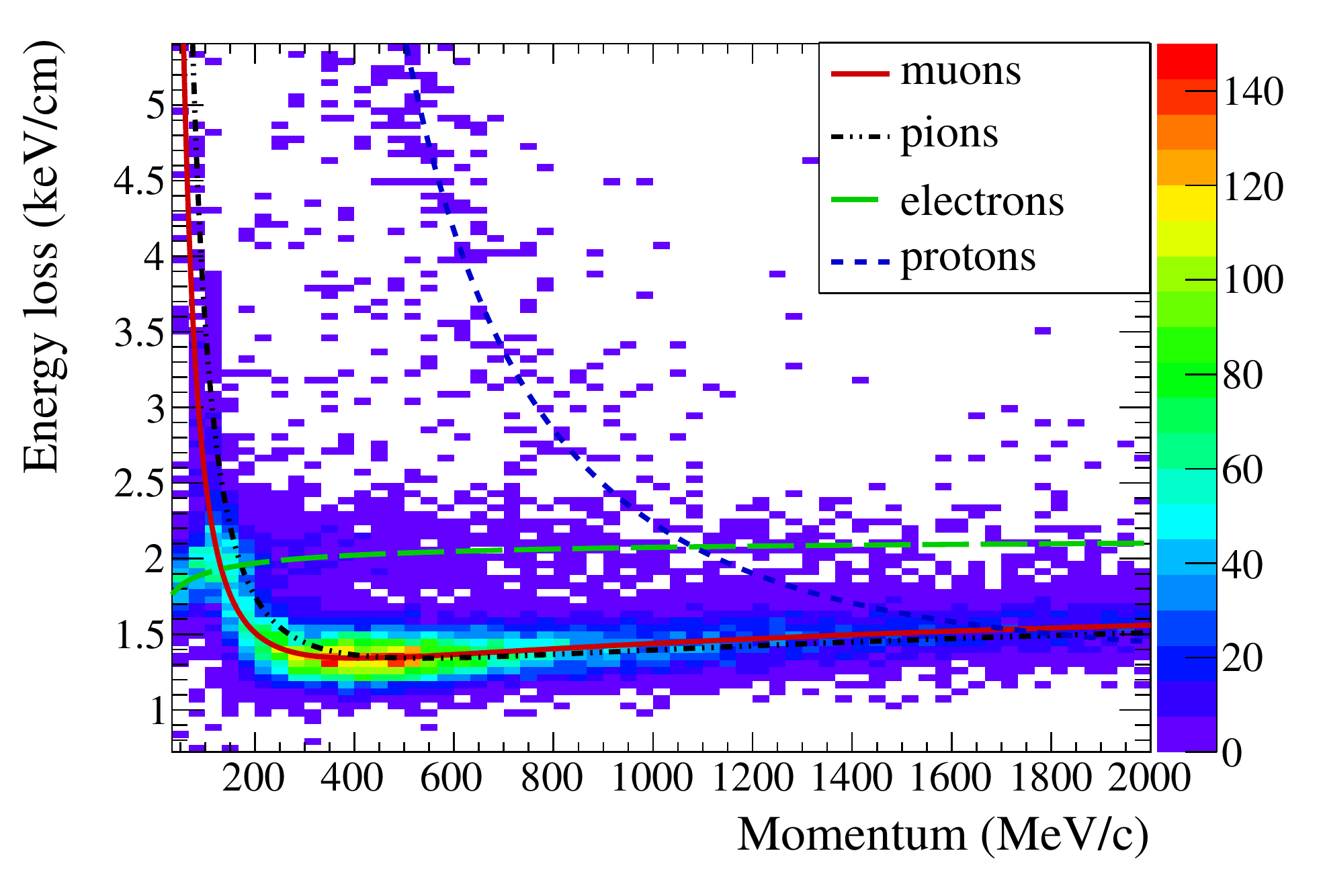}
\caption{\label{fig:dedxneg} TPC ionization energy loss per unit
  length as a function of the reconstructed momentum for negatively
  charged particles as measured by the TPCs (blocks) and the expected
  dependencies (curves).}
\end{figure}

\begin{figure}
\includegraphics[width=\linewidth]{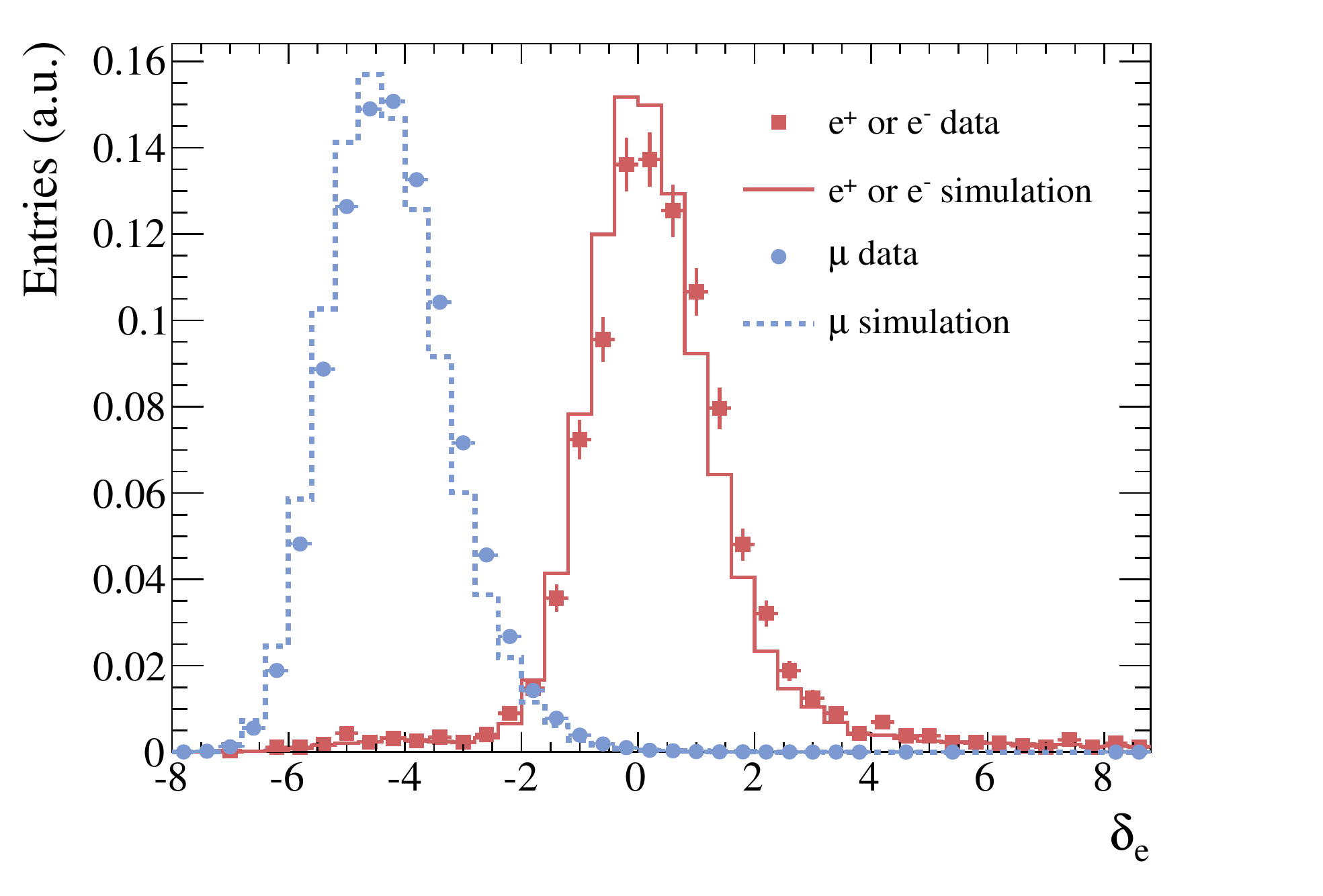}
\caption{\label{fig:tpcpull} TPC PID pulls in the electron hypothesis
  ($\delta_e$) for electrons or positrons coming from photon
  conversions and for muons in data (points) and in the simulation
  (lines).}
\end{figure}

The deposited energy as a function of the reconstructed momentum for
negatively charged particles starting in the FGD, compared with the
expected curves from the simulation is shown in
Fig.~\ref{fig:dedxneg}.  In the energy region of interest for T2K, the
ionization difference between electrons and muons is 30--40\% allowing
good separation between the two particles.  Fig.~\ref{fig:tpcpull}
shows the distribution of the pulls in two samples: the first is a
sample of muons which cross the detector whose selection is described
in Sect.~\ref{sec:mumisid}, and the second is a sample of electrons
and positrons selected as described in Sect.~\ref{sec:gammasel}.


\subsection{\label{sec:ecalpid} ECal reconstruction and PID performance}

Each ECal module has scintillator bars in two orientations. The
reconstruction is performed by forming two sets of 2D clusters, one for
each orientation, then combining them to form a 3D cluster. The 2D
objects are built by clustering together adjacent hits. If more than
one 2D object exists in a particular orientation, the choice of which
should be used in the 3D object is based on a likelihood statistic combining
time, position and charge information, aided by the extrapolation of in-time TPC tracks.

After an ECal cluster is reconstructed, PID statistics to classify the
cluster are calculated. In particular, $\mipem$ is a statistic to separate
electromagnetic showers and minimally ionizing tracks and is a likelihood ratio using characteristics that distinguish tracks
and showers:

\begin{itemize}
\item circularity: clusters due to tracks are expected to be long
and thin, while showers are expected to have a more spherical shape;

\item charge distribution: electromagnetic showers have a highly
  non-uniform charge distribution compared to a minimally ionizing
  track. The charge distribution is parameterized using the ratio of the
  second and first moments and the ratio of the highest charge to the
  lowest charge layer;

\item charge ratio between first quarter and last quarter of the
  track: it is expected to be one for minimally ionizing tracks which
  deposit energy uniformly, greater than one for electromagnetic showers
  and less than one for highly ionizing particles such as protons
  which deposit most of their energy at the end of the track.
\end{itemize}

\begin{figure} [htbp]
    \begin{center}
      \includegraphics[width=\linewidth]{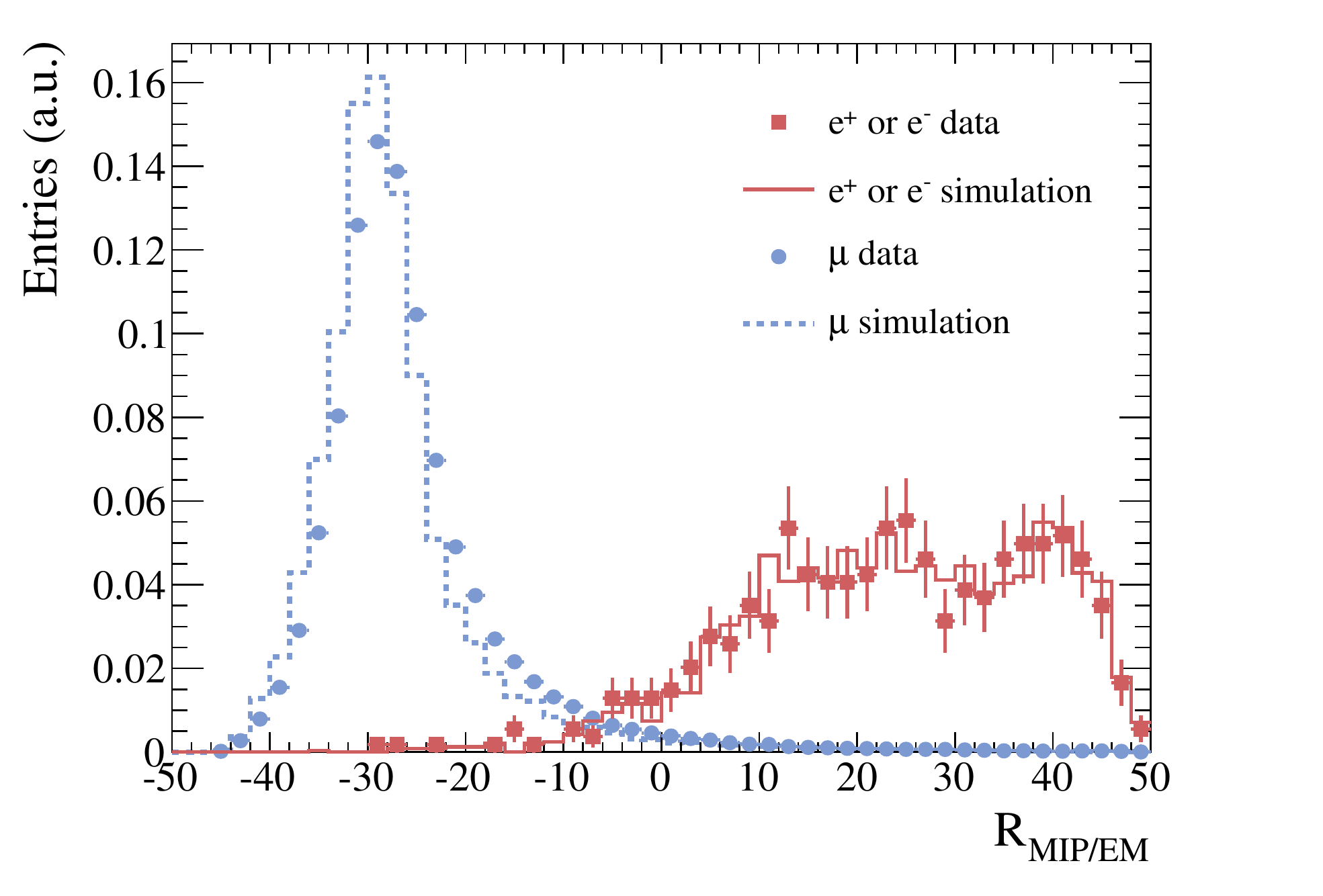}
      \caption{
           Distribution of $\mipem$ for electrons or positrons
            coming from photon conversions, and for muons in
            the downstream ECal in data (points) and in the simulation
            (lines).
        }
        \label{fig:ecal_pid_mip_em_distribution}
    \end{center}
\end{figure}

Samples of simulated electrons and muons 
are used to generate probability density functions (PDFs) that are used to construct the
likelihood ratio. Fig.~\ref{fig:ecal_pid_mip_em_distribution} shows
the $\mipem$ statistic in data and simulation for samples of \epem
from photon conversions and from crossing muons. 

The energy deposited
in the ECal (\emene) is used for particles with reconstructed momenta
in the TPC larger than 1~\gev/c, to discriminate between electrons and
muons. A charged particle that enters the ECal from the TPC
has momentum measured in the tracker and this can be compared to the
energy deposited in the ECal. Energy is reconstructed
under the hypothesis that the energy deposit is due to an
electromagnetic shower. A maximum likelihood fit for the shower energy
is constructed using the following variables:

\begin{itemize}
\item the total visible energy in the cluster: the total energy
  deposited into the scintillator is strongly correlated to the energy
  of the particle responsible for the EM shower and this parameter
  dominates the energy measurement in the ECal;
\item the RMS and the skewness of the deposited energy: these parameters provide additional information that refines the energy measurement.
\end{itemize}

The fit uses PDFs constructed from simulated photons at energies from
50 MeV to 25 GeV, with the majority of photons below 2 GeV. The energy
resolution for electrons at 1~\gev is approximately $10\%$. 
 
\section{\label{sec:nuesel} Selection of electron neutrino CC events at ND280}

\begin{figure}
\includegraphics[width=\linewidth]{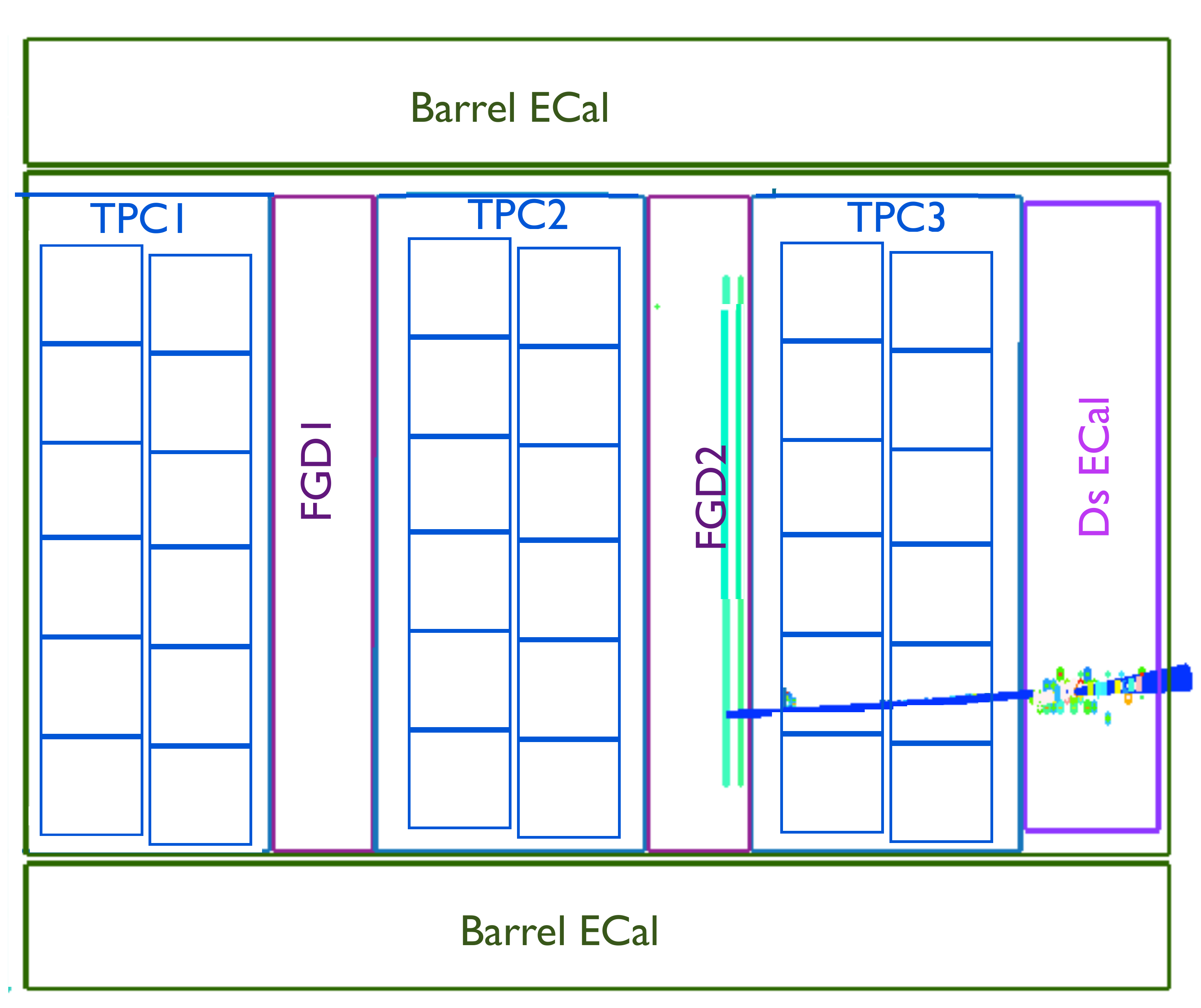}
\caption{\label{fig:nue_ev_disp}Side view of a \nue CCQE-like event in FGD2 with an electron-like track reconstructed in TPC3 and showering in the downstream ECal.}
\end{figure}

The signal events for this analysis are \nue CC interactions occurring in FGD1 or FGD2. Events in which there are electron-like tracks starting in either FGD are selected, and additional cuts are applied to reduce the contamination from photons converting into an \epem pair in an FGD. The events are then split into separate CCQE-like and CCnonQE-like samples. A typical \nue CC candidate selected in the analysis is shown in Fig.~\ref{fig:nue_ev_disp}.

After requiring a good beam spill and good ND280 data quality---all subdetectors were functioning correctly---the reconstructed objects in each spill are split into 8 time bunches (6 for Run I). For each bunch the highest momentum negatively charged track is selected as the lepton candidate. If this track does not start in the fiducial volume (FV) of one of the FGDs the event is rejected. The FGD fiducial volume is defined by removing the outer 48~mm at each edge in $x$ and $y$ (distance equivalent to five scintillator bars) and the front 21~mm (7~mm) at the begin of the FGD1 (FGD2), corresponding to the first $x$-$y$ ($x$) layer.

The track is also rejected if the reconstructed momentum is smaller than 200~MeV/c as that region is dominated by background from photon conversions. To ensure good TPC PID performance the selected track needs to have at least 36 reconstructed clusters in the TPC, corresponding to tracks crossing at least half of the TPC in the direction parallel to the beam. 

\begin{figure}
\includegraphics[width=\linewidth]{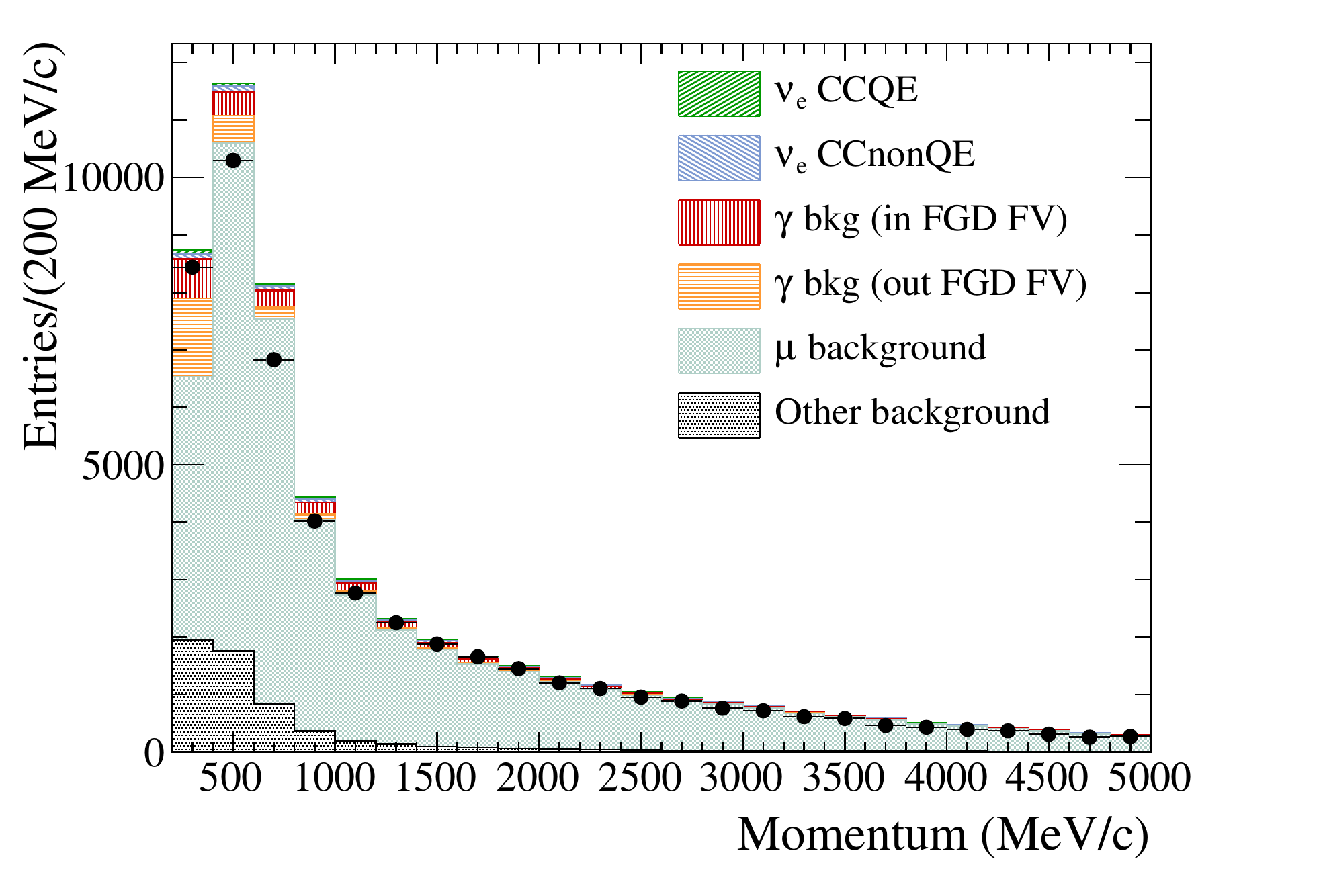}
\includegraphics[width=\linewidth]{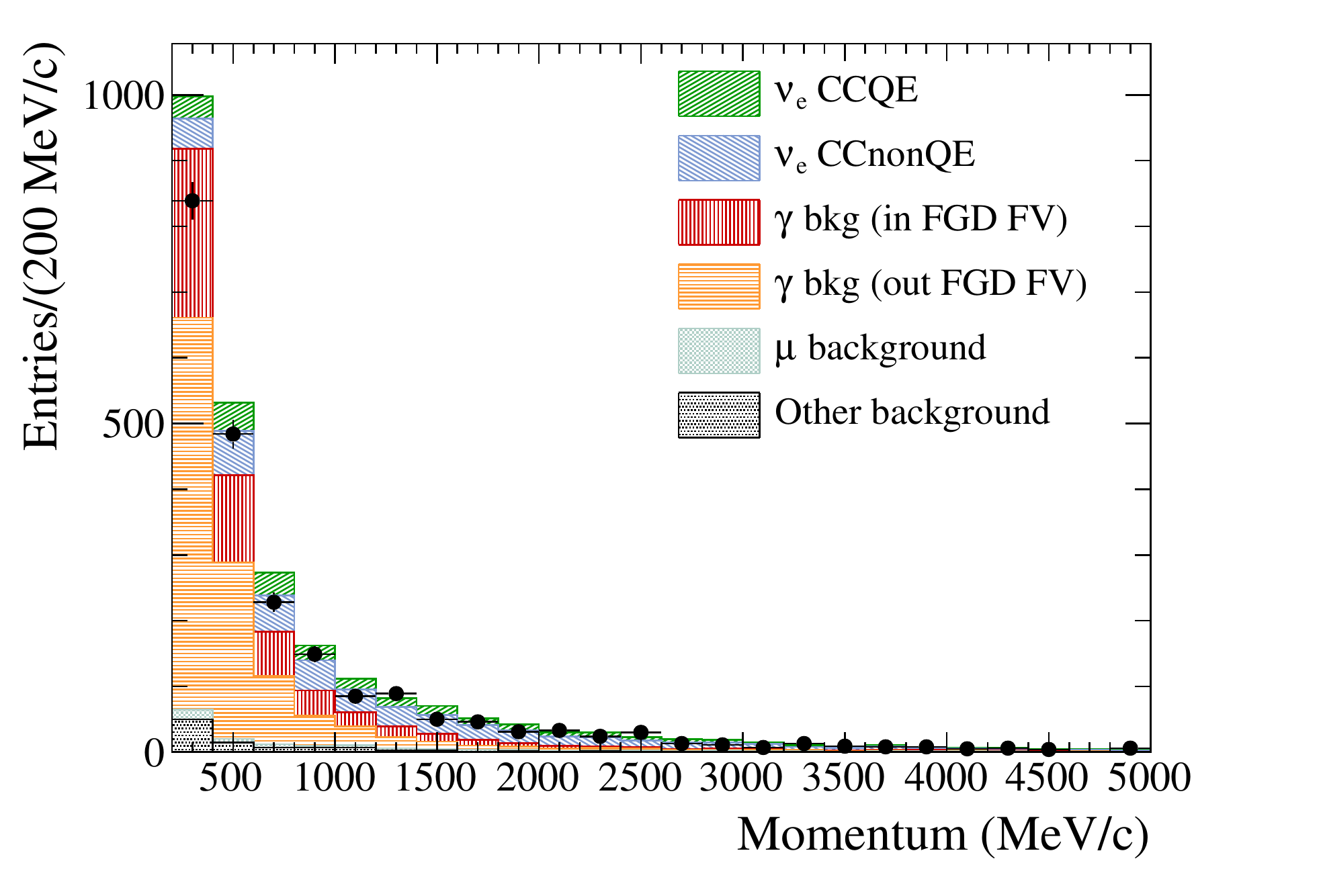}
\caption{\label{fig:nue_postpid_reac}Reconstructed electron momentum
  of events before (top) and after (bottom) the PID selection. The signal is divided into \nue producing CCQE or CCnonQE interactions. The background is divided into photon conversions produced by neutrino interaction inside or outside the FGD, misidentified muon background and other background (mainly pions and protons). The error on the points is the statistical error on the data.}
\end{figure}

Applying these criteria, 79\% of the tracks are expected to be muons and just 6.5\% electrons (see the inset in Fig.~\ref{fig:nue_postpid_reac}). To select electrons, the TPC and ECal PID capabilities are combined. The PID criteria applied depend upon which sub-detectors are used for the track reconstruction:
\begin{itemize}
\item if the electron candidate does not enter the ECal, the energy loss in the TPC is required to be electron-like ($-1<\delta_{e}<2$), not muon-like ($|\delta_{\mu}|>2.5$) and not pion-like ($|\delta_{\pi}|>2$). This selection is also used for all tracks with reconstructed momentum in the TPC below 300~MeV/c as the ECal PID is not optimized for such low energy particles;
\item for tracks entering the ECal, the TPC PID is relaxed, only requiring an electron-like track ($-2<\delta_{e}<2.5$). The ECal particle identification criteria depend on the momentum of the track as it enters the ECal module. For tracks with a momentum greater than 1~GeV/c, the energy deposited in the ECal module is used to separate electromagnetic showers from minimum ionizing particles. Tracks are required to have $\emene>1100$~MeV. For lower-momentum particles, the multi-variate analysis quantity \mipem is used. These tracks must have $\mipem>0$.
\end{itemize}

Tab.~\ref{tab:nue_paths} shows the performance of the different PID cuts, and highlights the effectiveness of combining the TPC and ECal information.

The momentum distribution of the particles passing the PID cuts is shown in Fig.~\ref{fig:nue_postpid_reac}. 99.9\% of muons are rejected by the PID cuts, and the sample is 92\% pure in electrons. Although a high-purity sample of electrons has been selected, 65\% of the tracks arise from $\gamma\rightarrow\epem$ conversions in the FGD, and only 27\% are from \nue CC interactions. The majority of the photons come from neutrino interactions upstream of the FGD in which the conversion occurred.

\begin{table}
  \caption{\label{tab:nue_paths} Fraction of electrons entering each PID branch, and efficiency and purity of the PID selection.}
\begin{ruledtabular}
 \begin{tabular}{ccccc}
& \multicolumn{2}{c}{FGD1 vertices} & \multicolumn{2}{c}{FGD2 vertices} \\
\hline
\multirow{2}{*}{Category}  & \multicolumn{1}{c}{events} & \multicolumn{1}{c}{eff. (\%)} & \multicolumn{1}{c}{events} & \multicolumn{1}{c}{eff. (\%)} \\
 &   \multicolumn{1}{c}{(\%)} & \multicolumn{1}{c}{[pur. (\%)]} &  \multicolumn{1}{c}{(\%)} & \multicolumn{1}{c}{[pur. (\%)]} \\
\hline 
 \multirow{2}{*}{TPC only} & \multirow{2}{*}{45.4} & 56.6 & \multirow{2}{*}{34.1} & 53.1 \\
 & & [92.6] & & [90.9] \\

 \hline
\multirow{2}{*}{TPC+DsECal}& \multirow{2}{*}{32.0} & 82.6 & \multirow{2}{*}{59.0} & 89.1 \\
 & & [97.8] & & [93.8] \\
\hline
TPC+Barrel& \multirow{2}{*}{22.6} & 86.1 & \multirow{2}{*}{6.9} & 88.6 \\
ECal & & [91.4] & & [86.5] \\

 \end{tabular}
\end{ruledtabular}
\end{table}

To reduce the contamination from these photons, veto cuts are applied to require no reconstructed tracks in the P0D, TPC or Barrel ECal in the same bunch, starting more than 100~mm upstream of the initial position of the electron candidate. 

An additional cut in the selection removes electrons that are part of an \epem pair. The event is rejected if there is a positive track which is electron-like ($|\delta_{e}|<3$), starts within 100~mm of the electron candidate, and if the \epem pair has an invariant mass of less than 100~MeV/c${}^2$.
These cuts reduce the $\gamma\rightarrow\epem$ contamination from 65\% to 30\%.

To further improve the \nue purity, additional selections are applied and the sample is separated into CCQE-like and CCnonQE-like categories. The first mainly contains \nue CCQE interactions while the latter is dominated by \nue CC interactions producing pions in the final state.

The CCQE-like selection requires the absence of other tracks in the TPC, except the electron candidate itself. If the electron candidate starts in FGD1 then there must be no isolated reconstructed tracks in FGD1 and no Michel electrons coming from pion decays (identified as delayed hits in FGD1). These two requirements do not apply to events in FGD2 as the lower number of scintillator layers reduces the ability to reconstruct tracks and identify delayed hits.

If the electron candidate starts in FGD2 there must be no activity in the ECal, except that caused by the electron candidate. This cut is only applied to events in FGD2, as electrons from FGD1 can shower in FGD2 and can cause additional ECal activity not associated with the original electron candidate track. 

The CCnonQE-like selection requires the presence of at least one other track which starts close to the electron candidate (within 50~mm). As in the CCQE-like selection, only FGD-TPC tracks are considered for FGD2 events, whereas FGD-only tracks are also considered for FGD1 events. For FGD1 events the presence of a Michel electron in the FGD is used to tag CCnonQE-like candidates.

\begin{figure}
\includegraphics[width=\linewidth]{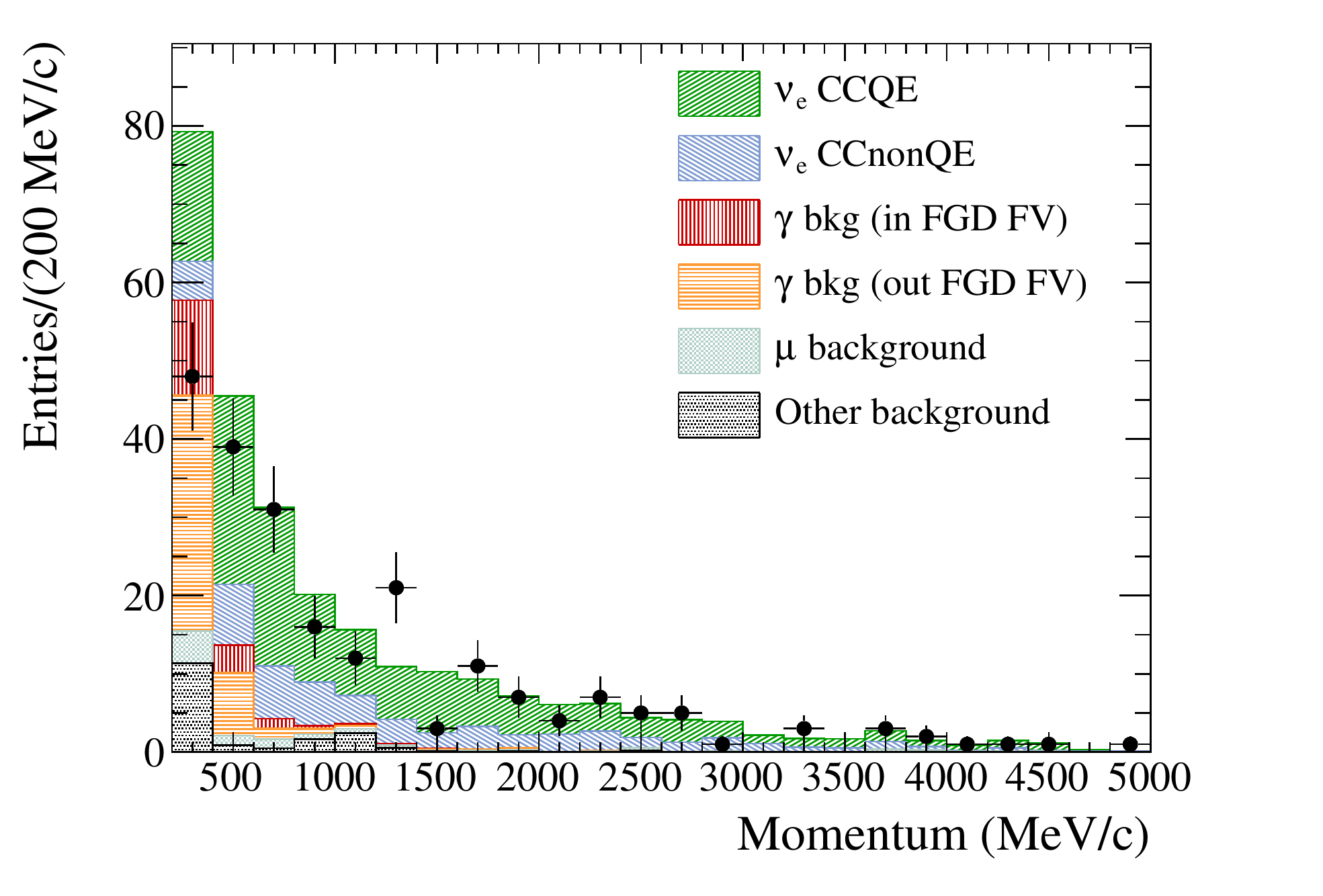}
\caption{\label{fig:nue_ccqe_mom}Reconstructed electron momentum of events selected in the CCQE-like selection. The error on the points is the statistical error on the data, and the simulation is divided into the same categories as Fig.~\ref{fig:nue_postpid_reac}.}
\end{figure}

\begin{figure}
\includegraphics[width=\linewidth]{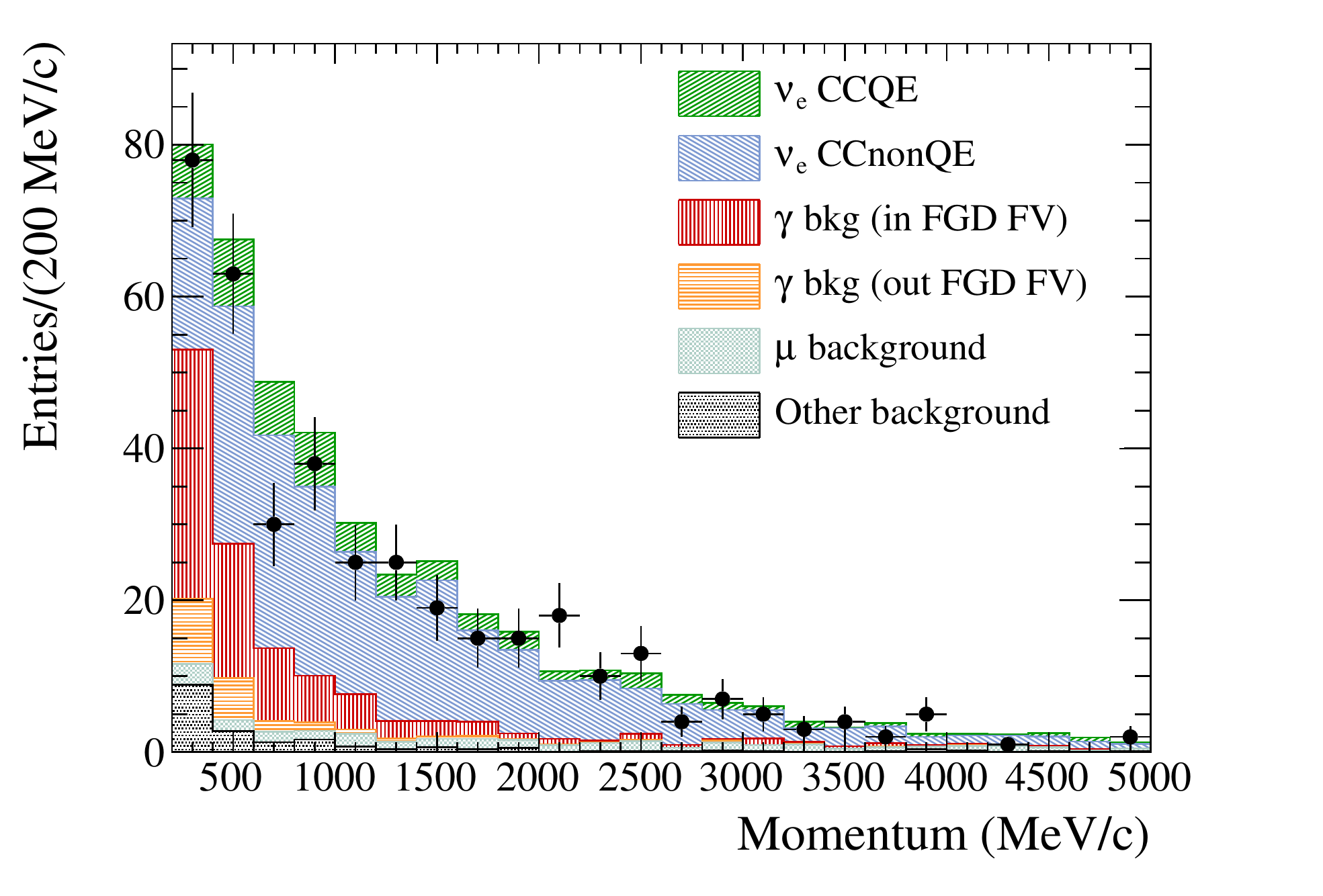}
\caption{\label{fig:nue_ccnqe_mom}Reconstructed electron momentum of events selected in the CCnonQE-like selection. The error on the points is the statistical error on the data, and the simulation is divided into the same categories as Fig.~\ref{fig:nue_postpid_reac}.}
\end{figure}

\begin{figure}
\includegraphics[width=\linewidth]{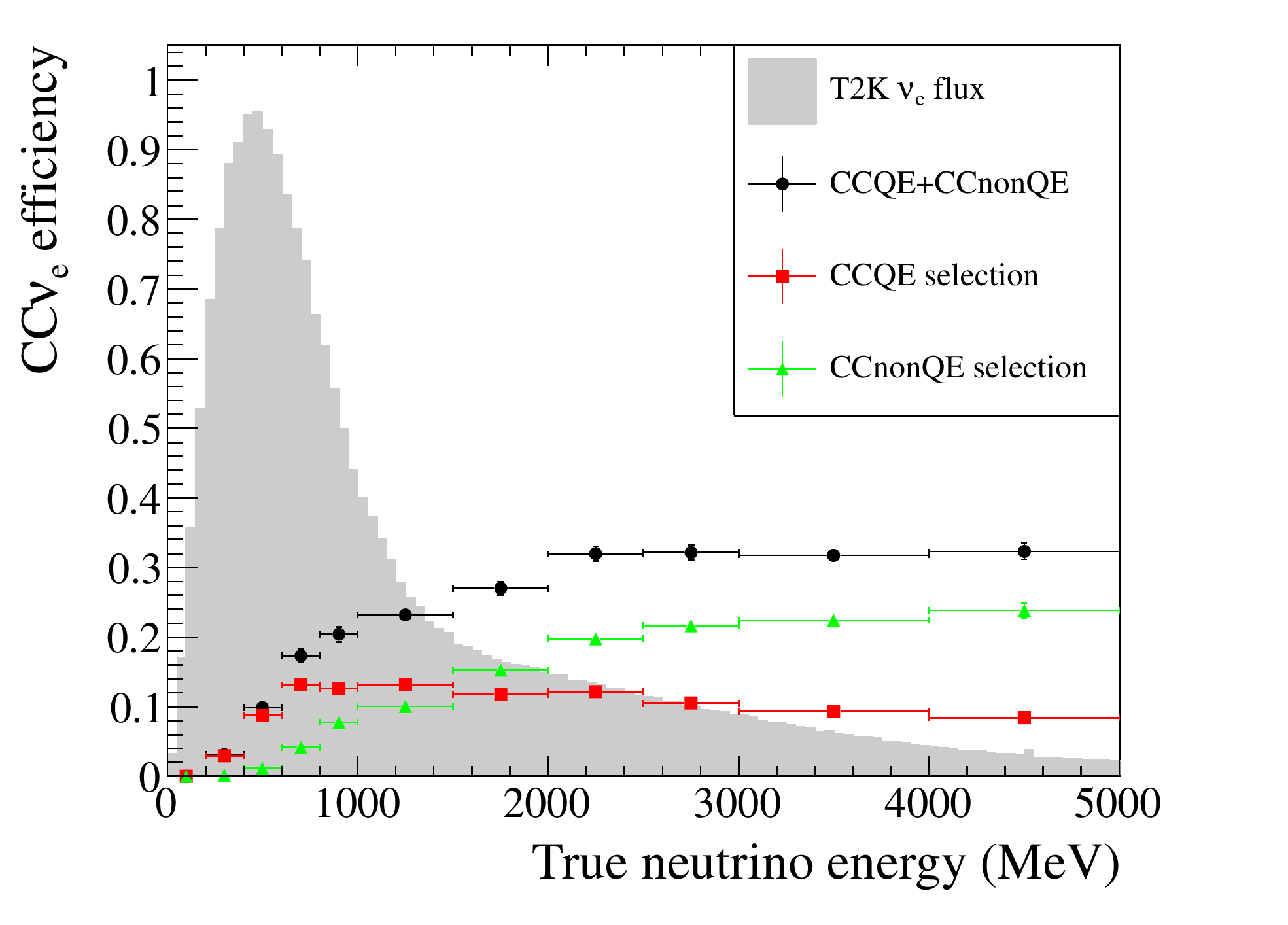}
\caption{\label{fig:nue_eff_inc_e}Efficiency of selecting \nue CC interactions as a function of true neutrino energy, and the predicted \nue flux at ND280.}
\end{figure}

The final CCQE-like and CCnonQE-like selections are shown in Fig.~\ref{fig:nue_ccqe_mom} and Fig.~\ref{fig:nue_ccnqe_mom} respectively. The overall efficiency of selecting \nue CC interactions is 26\%, and the efficiency of the selections as a function of \nue energy is shown in Fig.~\ref{fig:nue_eff_inc_e}. The purity of the selections and the predicted number of selected events are shown in Tab.~\ref{tab:nue_purity}.

\begin{table}
  \caption{\label{tab:nue_purity} Fractions, expected, and observed number of events for the CCQE and CCnonQE selections.}
\begin{ruledtabular}
 \begin{tabular}{ldddd}
 \multirow{2}{*}{Category} & \multicolumn{2}{c}{CCQE selection} & \multicolumn{2}{c}{CCnonQE selection} \\
  & \multicolumn{1}{c}{Fraction (\%)} & \multicolumn{1}{c}{Events} & \multicolumn{1}{c}{Fraction (\%)} & \multicolumn{1}{c}{Events} \\
\hline 
 \nue CCQE & 48.2 & 132.6 & 12.7 & 56.8 \\
 \nue CCnonQE & 19.6 & 54.1 & 52.8 & 234.7 \\
 $\gamma$ bkg (in FGD) & 6.4 & 17.8 & 19.2 & 85.3 \\
 $\gamma$ bkg (out FGD) & 15.0 & 41.4 & 4.5 & 19.9 \\
 $\mu$ background & 4.0 & 10.9 & 6.2 & 27.6\\
 Other background & 6.8 & 18.8 & 4.6 & 20.6\\ 
\hline
Total simulation & 100.0 & 275.6 & 100.0 & 444.9 \\
\hline
\bf{Data} & & 225 &  & 392 \\ 
\end{tabular}
\end{ruledtabular}
\end{table}

\section{\label{sec:background} Control of the backgrounds }
The selection of \nue CC interactions is designed to reject two large
backgrounds. The first one is due to the muons produced in $\nu_{\mu}$
CC interactions that
are the dominant component of the T2K beam. This component is rejected using the PID
capabilities of ND280. The second background is due to the
conversions of photons
in the FGD producing electrons in the TPC and it cannot be rejected
using PID algorithms.

For the muon background, the combined PID of the TPC and ECal is vital
to reject 99.9\% of the muons. Such a large muon rejection power has been verified using
a clean, data driven sample of muons, as described in
Sect.~\ref{sec:mumisid} below.
The photon background is constrained using a selection of
photon conversions in the FGD in which both the electron and the
positron are reconstructed in the TPC, as described in
Sect.~\ref{sec:gammasel} below.

\subsection{\label{sec:mumisid} Muon misidentification}

\begin{figure}
\includegraphics[width=\linewidth]{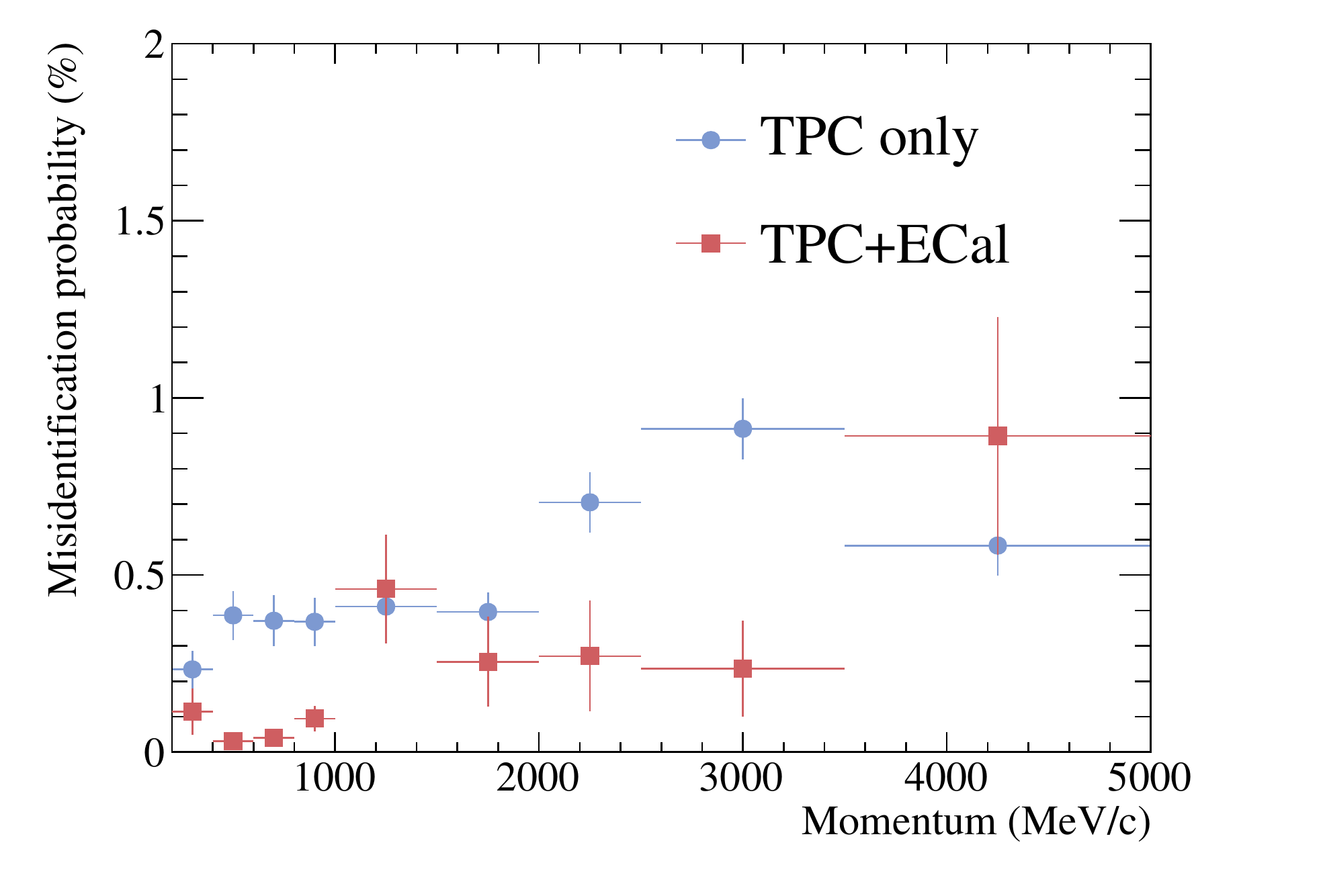}
\caption{\label{fig:misid_muon_prob} 
Muon misidentification probability as a function of the muon momentum
estimated using a sample of through going muons for the case in which the TPC PID
only is used and for the case in which TPC and ECal PID are combined.}
\end{figure}

A data-driven study has been carried out to confirm the 
muon rejection power expected by simulation. A clean sample of muons, produced by neutrino
interactions in the sand or in the concrete walls of the ND280 pit, is
selected. The selection is done by requiring
one and only one track in a bunch with negative charge crossing all the
3 TPCs and starting at the upstream edge of the P0D. The TPC PID of the selected track must be compatible with a
muon 
in the TPC upstream of the first FGD (TPC1). This requirement does
not bias the sample since the TPC1 PID is not used in the analysis.
Once a clean sample of muons is selected from the data, the muon
misidentification probability is computed as the ratio between the
number of tracks passing the PID selection and all the selected tracks. The same PID selections
described in Sect.~\ref{sec:nuesel} for the cases with and without ECal information are
used. In Fig.~\ref{fig:misid_muon_prob} the misidentification
probability as a function of the track momentum measured in the TPC is shown.
The muon misidentification probability is below 1\% for all the
momenta and is much
smaller if the TPC and the ECal PID are combined. Compatible results are
obtained using simulated data confirming that the PID
performances of ND280 detectors are well understood and well
reproduced in the simulation.

\subsection{\label{sec:gammasel} Photon selection}
The background, especially at low momentum, is dominated by electrons coming from
photon conversions. Those electrons are background to the \nue CC
analysis as they typically come from $\nu_{\mu}$ interactions that occur inside or outside
the FGD producing a $\pi^0$ in the final state which immediately decays into two photons.
One of the two photons then converts inside the FGD producing an
\epem pair. If the positron is not reconstructed in the
TPC the event is
topologically equivalent to a $\nu_e$ CC interaction. 

\begin{figure}
\includegraphics[width=\linewidth]{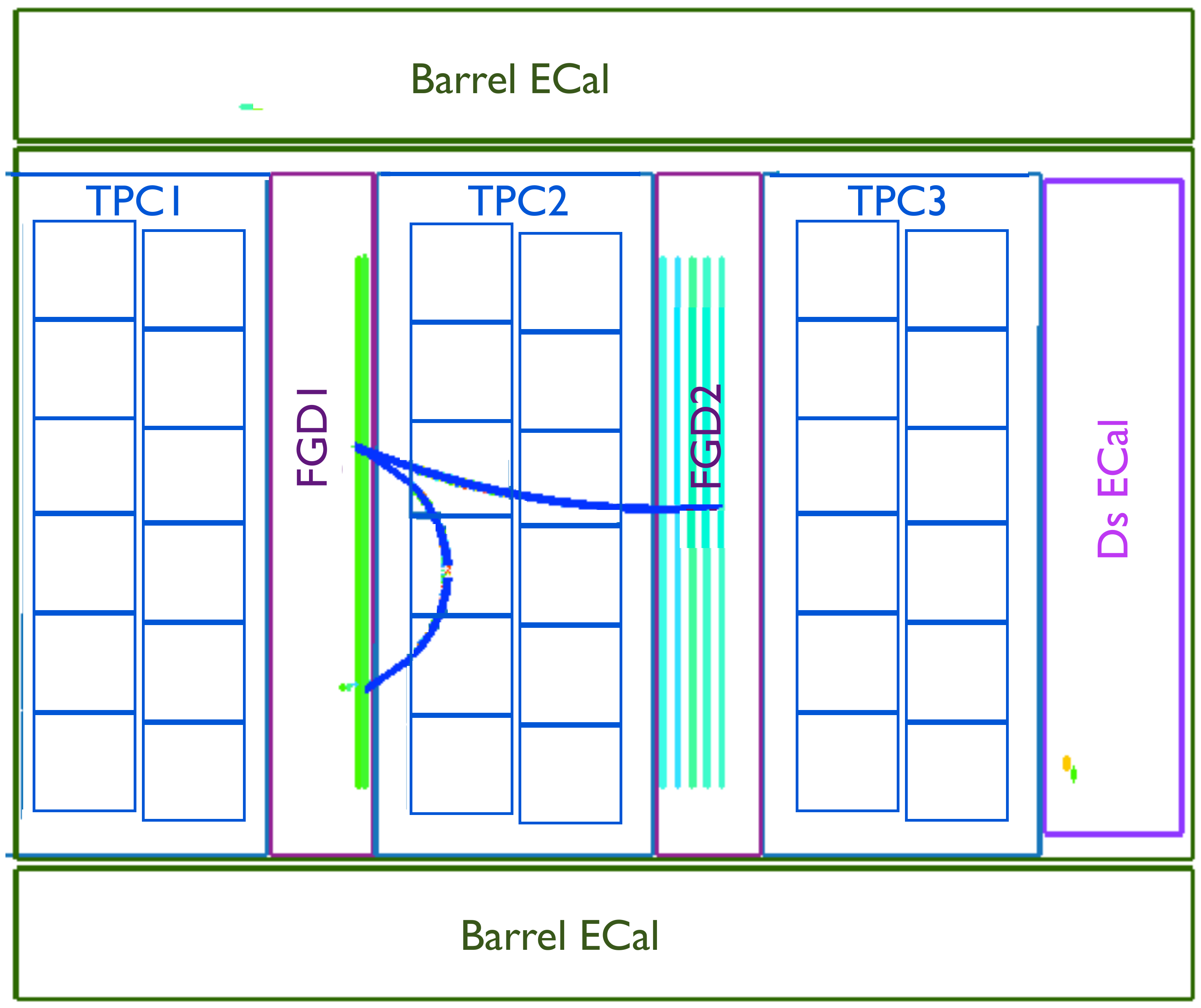}
\caption{\label{fig:gm_ev_disp}Side view of a photon conversion in
  the FGD1 with an \epem pair reconstructed in TPC2.}
\end{figure}

This background can be estimated using a selection of photon
conversions in which both the electron and the positron are
reconstructed. A typical example of  a photon conversion with the
\epem pair reconstructed in the TPC is shown in
Fig.~\ref{fig:gm_ev_disp}.

To select these events, two tracks are required to
start in the FGD fiducial volume and to have
opposite charge. The same data quality criteria described in
Sect.~\ref{sec:nuesel} are also required. 

Both tracks have to be electron-like ($|\delta_e|<2$ for the
negatively charged track and $|\delta_e|<3$ for the positively charged track),
the distance of the starting point of the two tracks
is required to be within 100~mm, and the reconstructed invariant mass
of the pair has to be smaller than 50~MeV/c${}^2$. 

\begin{figure} [h]
\begin{center} 
\includegraphics[width=\linewidth]{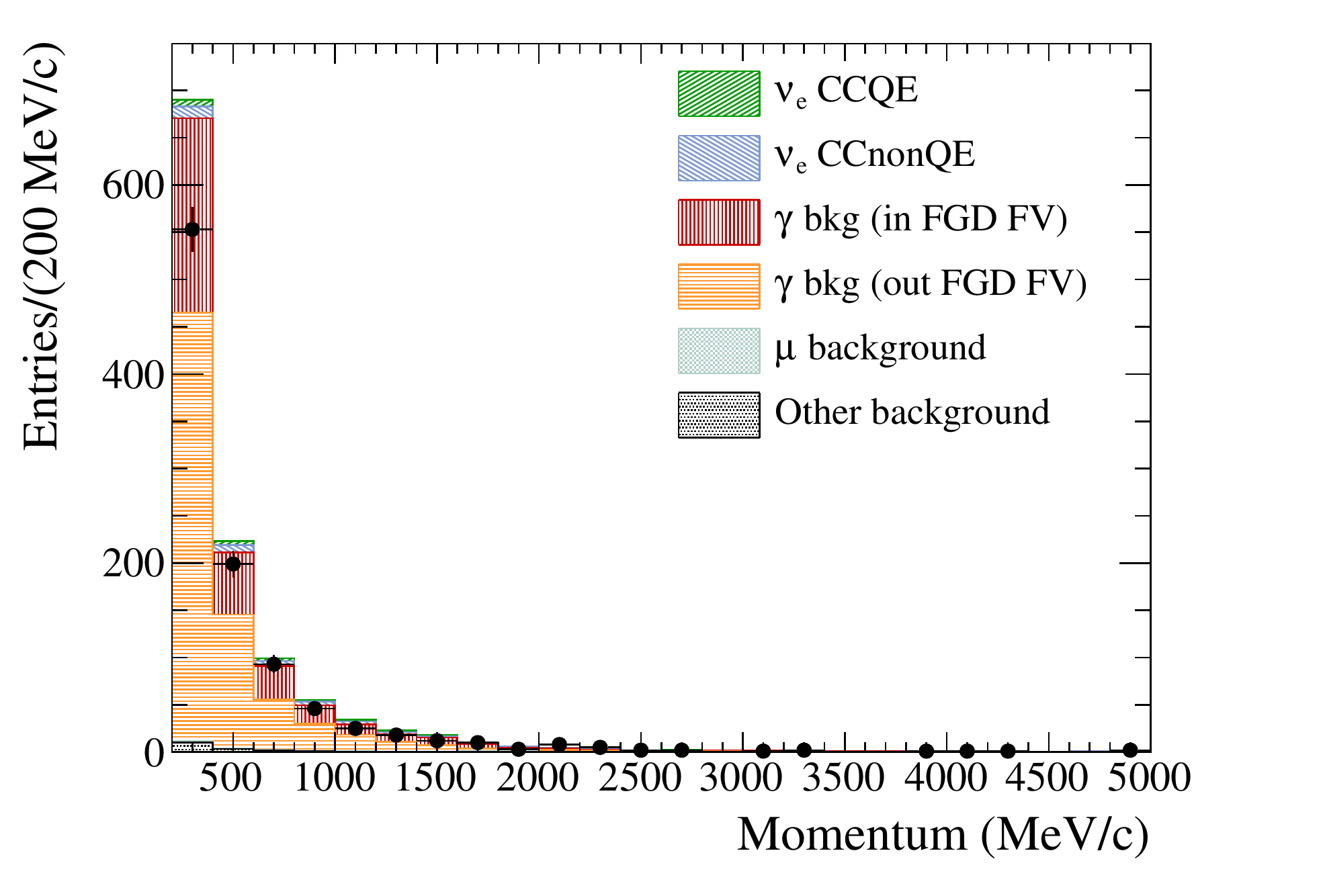}
\end{center}
\caption{Reconstructed electron momentum of events selected in
  the photon selection. The error on the points is the statistical error on the data, and the simulation is divided into the same categories as Fig.~\ref{fig:nue_postpid_reac}.}
\label{gm:sel_fam}
\end{figure}

The most powerful requirement among those is the invariant mass cut that alone
is able to select a sample with a 90\% purity in electrons.
After applying all the criteria a sample with an electron purity of
$99\%$ is selected.
The momentum of the electrons in the selected events is shown in
Fig.~\ref{gm:sel_fam}. The purity of photon conversions in the sample
is $92\%$ with the remaining events mainly coming from \nue
interactions in which the electron showers in the FGD and produces
a positron in the TPC.
The efficiency of this selection with respect to the total number of
photons converting in the FGD is 12\%.

The purpose of this selection is to estimate the number of electrons
coming from photon
conversions entering the $\nu_e$ CC selection. In order to do this it is necessary
to ensure that the characteristics of the events selected in the photon selection and
in the photon background in the \nue CC selection are similar. Specifically
they need to have the same origin, they need to be produced in the
same type of neutrino
interactions, and the selected lepton in the two cases needs to cover the same phase
space. Tab.~\ref{tab:pur_pid} shows the neutrino interactions
contributing to the two samples and the fraction of neutrinos
interacting inside or
outside the FGD. The fractions of neutrino
interactions are similar while the photon selection has more events coming from outside of the FGD. This difference is due to the
different geometrical acceptance of the two samples because the photon
selection requires both tracks to enter the TPC while the \nue CC
selection requires only one TPC track. 

The lepton momentum and
angle, and the neutrino energy are in reasonable agreement
between the two selections. The strategy used to constrain the background
in the extraction of the beam \nue component will be described in Sect.~\ref{sec:fit}.

\begin{table}
\caption{\label{tab:pur_pid} Fractions (in \%) of the different
  interaction types and of the production point for the events
  selected in the photon selection and in the photon background to the \nue
  CC selection.}
\begin{ruledtabular}
 \begin{tabular}{ldd}
Interaction & \multicolumn{1}{c}{$\gamma$ background} &
\multicolumn{1}{c}{$\gamma$ selection} \\ 
(Production point) & \multicolumn{1}{c}{in \nue CC selection} & \\ \hline
CCQE & 4.3 & 4.1\\
CC$1\pi$ & 14.2 & 11.5\\
CC Coherent & 0.5 & 0.4\\
CC other & 43.5 & 41.7 \\
NC$1\pi^0$ & 8.6 & 10.9\\
NC other & 28.8 & 31.4\\
\hline
Inside FGD FV & 57.0 & 30.6\\
Outside FGD FV & 43.0 & 69.4\\ 
\end{tabular}
\end{ruledtabular}
\end{table}

\section{\label{sec:systematics} Systematic uncertainties }

The systematic uncertainties in the measurement of the beam \nue component are separated into three main categories: detector performance, external backgrounds, and neutrino flux and cross section uncertainties.
The systematic uncertainties described in this section are used as prior constraints in the fit to extract the beam \nue component that will be described in Sect.~\ref{sec:fit}. 

\subsection{\label{sec:detsyst} Detector systematic uncertainties}
The detector systematic uncertainties are computed for each subdetector used in this analysis: the TPCs, the FGDs and the different ECal modules. Systematic effects related to neutrino interactions outside ND280 producing particles entering the detector are also considered in this class of systematics.

To determine the effect of the detector systematic uncertainties on the analysis 1000 toy experiments have been performed. Each toy experiment has a set of detector systematic parameter values drawn from Gaussian distributions. The 1000 toy experiments are used to evaluate the covariance of the number of events in bins of reconstructed electron momentum and neutrino flavor for each selection. Seven bins are included for the CCQE-like and CCnonQE-like selection and 3 bins for the $\gamma$ selection.

\subsubsection{TPC systematic uncertainties}

TPC systematic uncertainties are divided into three classes: selection efficiency, momentum resolution, and particle identification.

The efficiency systematic uncertainty arises from the cluster finding, track finding and charge assignment. It is evaluated using events with a single particle passing through multiple TPCs, to check for the presence of a reconstructed track with the correct charge assignment. The TPC reconstruction efficiency is determined to be $(99.8^{+0.2}_{-0.4})\%$ and the charge misassignment probability is below 1\% for tracks with momenta less than 5~GeV/c.

Momentum reconstruction is affected by non-uniformity of the magnetic
field and the overall magnetic field strength. The field inside the
magnet has been measured with a Hall probe and non-uniformities are
checked with photoelectrons produced by shining a laser at the central
cathode of the TPC that has small aluminium dots on it. Uncertainty in
the overall magnetic field strength leads to an uncertainty on the
momentum scale of 0.6\%. An additional source of systematic
uncertainty is the momentum resolution that has been determined using
tracks crossing multiple TPCs and comparing their reconstructed
momenta. The inverse momentum resolution is found to be  significantly
better in simulations than in data, which could be due to
non-uniformity of the electric field. These non-uniformities depend on
the drift distance and cancel out for tracks close to the cathode. For
this reason they are not observed in the analysis of photoelectrons produced by the
laser. At momenta larger than 1.4~\gev/c, the inverse momentum resolution is ($30\pm10)\%$ larger in data than in simulation. 

Systematic uncertainties on the TPC particle identification are computed using high-purity samples of electrons, muons and protons. By definition the electron sample should give a $\delta_e$ distribution that is Gaussian with mean 0 and width 1. The simulation and the data have a difference on the mean of the pull distribution of ($-0.12\pm0.12$), and a scale of the width of ($1.02\pm0.07$). These are converted into systematic uncertainties on the energy loss for each true electron track, and all the TPC PID pulls are recomputed. Similarly, the $\delta_p$ distribution of a proton sample is used to determine the energy loss systematic uncertainties of true protons. The $\delta_{\mu}$ distribution of the through going muon sample is used for both muons and pions, as their masses are similar. 

\subsubsection{FGD systematic uncertainties}

The systematic uncertainties related to the FGDs arise from potential mismodelling of the track-finding efficiency, the efficiency with which TPC and FGD tracks are matched, the Michel electron tagging efficiency, secondary pion interactions, and the FGD mass. 

The efficiency with which FGD tracks are reconstructed is computed for FGD1, as the analysis uses isolated FGD reconstruction for FGD1 only. To determine the efficiency, samples of stopping proton-like tracks in FGD1 are used and differences between the data and the simulation are evaluated for different angles, being 3\% for forward-going tracks and rising to 21\% for high angle tracks. Additional studies have been done for multiple tracks in the FGD by using hybrid samples in which simulated particles (pions or protons) are injected into data and simulated events finding differences between data and simulation of 3\% (4\%) for pions (protons). 


The TPC-FGD matching efficiency is studied using a sample of through-going particles, in which the presence of a track in the TPC upstream and downstream implies that a track should be seen in the FGD. The efficiency is found to be 99.9\%. The matching efficiency has also been checked for electrons using samples of photon conversions in the FGD in which the \epem tracks were not required to start in the FGD. 

The Michel electron tagging efficiency is studied using a sample of cosmic rays which stop in FGD1. The particles must have a range compatible with being a muon with the momentum as reconstructed in the TPC. The Michel tagging efficiency is found to be about 60\%, depending on the beam bunch of the primary neutrino interaction.

There is an uncertainty in the modeling of pion reinteractions (where a pion created in an interaction is ejected from the nucleus and interacts with another nucleus in the detector).
Differences between external pion interaction data~\cite{dePerio:2011zz} and GEANT4 simulation are evaluated. The external data are extrapolated to cover the whole momentum region relevant for T2K and a correction weight is calculated for each event, along with an uncertainty based on the error of the extrapolated data. This systematic uncertainty can only migrate events between the CCQE-like and the CCnonQE-like selections and the effect on this analysis is found to be negligible.

The uncertainty on the mass of the FGDs---and thus on the number of target nucleons---is computed using the known density and size of the individual components, and their uncertainties. The overall uncertainty on the FGD mass is 0.67\%.

\subsubsection{ECal systematic uncertainties}

ECal systematic uncertainties are computed for the particle identification, the energy resolution and scale, and the efficiency with which ECal objects are reconstructed and matched to TPC tracks.

Uncertainties in \mipem are calculated using high-purity samples of electrons and muons. The efficiency of the $\mipem>0$ requirement is calculated for data and for the simulation, and the difference is taken as the systematic uncertainty. The systematic uncertainty for selecting electrons is 2.1\% in the DsECal and 2.9\% in the Barrel, and the uncertainty for rejecting muons is 0.6\% in the DsECal and 0.7\% in the Barrel ECal.

\emene systematic uncertainties are computed using both a high-purity sample of electrons from T2K data and test beam data taken at CERN before the DsECal was shipped to Japan. The reconstructed \emene is compared to the known momentum, either as measured by the TPC for the in-situ sample, or as defined in the test beam. The uncertainty on the energy scale is 6\% and the uncertainty on the energy resolution is 15\%.

A combined systematic uncertainty is computed for the efficiency of reconstructing an ECal object and matching it to a TPC track. A high-purity sample of electrons that appear to enter the ECal (by extrapolating the TPC track) is used. The uncertainty is 1.6\% for tracks entering the DsECal, and 3.4\% for tracks entering the Barrel ECal.
 
\subsection{\label{sec:outfvsyst}External background systematic uncertainties}

\subsubsection{Systematic uncertainties from neutrino interactions outside the FGD FV}
In the \nue CC selection there is a large background component coming from $\nu_{\mu}$ interactions outside of the FGD. Differences between data and simulation in this component might arise from mismodelling of the $\pi^0$ production and in the description of the material and efficiencies outside of the FGD. The fit to the ND280 $\nu_{\mu}$ CC samples cannot constrain this component as $<5\%$ of the events selected in that analysis come from outside of the FGD.


In this analysis, the majority of the selected events arising from neutrino interactions outside of the FGD are due to photon conversion in the FGD and this component is well measured by the photon selection. The remainder of this background is due to other charged particles (pions or protons) and this component cannot be measured by the photon selection. 
The systematic uncertainty has been evaluated with control samples in which different sources of external backgrounds
producing tracks reconstructed as starting inside the FGD were
selected. The systematic uncertainty is set to $30\%$ for each of the
two components. 

\subsubsection{Systematic uncertainties from neutrino interactions outside ND280}
The standard ND280 simulation only includes neutrino interactions inside the magnet volume. Neutrino interactions outside this volume can affect the analysis in two different ways. First, the particles produced in these interactions can enter the ND280 volume, interact in an FGD, and be selected as signal. A separate simulation of neutrino interactions outside the ND280 magnet is used to estimate the expected number of extra signal events and is found to be negligible.
Particles produced outside ND280 can also enter in one of the ND280 sub-detectors in the same beam bunch in which a true \nue signal interaction occurs in an FGD. These particles can then cause an event to be vetoed, reducing the selection efficiency in data. The probability of this occurring is calculated separately for each veto cut and for each T2K run, as it depends on the beam intensity. The largest correction is $(2.3\pm0.5)\%$ for the ECal veto cut in Run IV.

\subsection{\label{sec:physsyst} Flux and cross section systematic uncertainties}

For the T2K oscillation analyses, the flux and the cross section systematic uncertainties described in Sect.~\ref{sec:t2k} are evaluated using a selection of $\nu_{\mu}$ CC interactions in ND280~\cite{Abe:2013hdq}. Events with a muon candidate produced in FGD1 and entering the downstream TPC are divided into the following samples:

\begin{itemize}
\item CC0$\pi$ sample in which there are no tracks compatible with a pion. Most of the events arise from CCQE interactions;
\item CC1$\pi$ sample in which there is one reconstructed track (in the TPC or in the FGD) compatible with a pion or there is a delayed energy deposit consistent with a stopping pion. It is mainly composed of CC interactions with resonant pion production;
\item CC-other sample, for all other topologies. Most events are from deep inelastic scattering and multi-pion production. 
\end{itemize}

These three samples are fit to evaluate the \num flux and cross section parameters and their uncertainties. The oscillation analyses also use these samples to evaluate the \nue flux parameters, which is possible because \num and \nue arise from the same parent particles. The cross sections for \nue and \num interactions are assumed to be the same.

\begin{figure}
\begin{center}
\includegraphics[width=\linewidth]{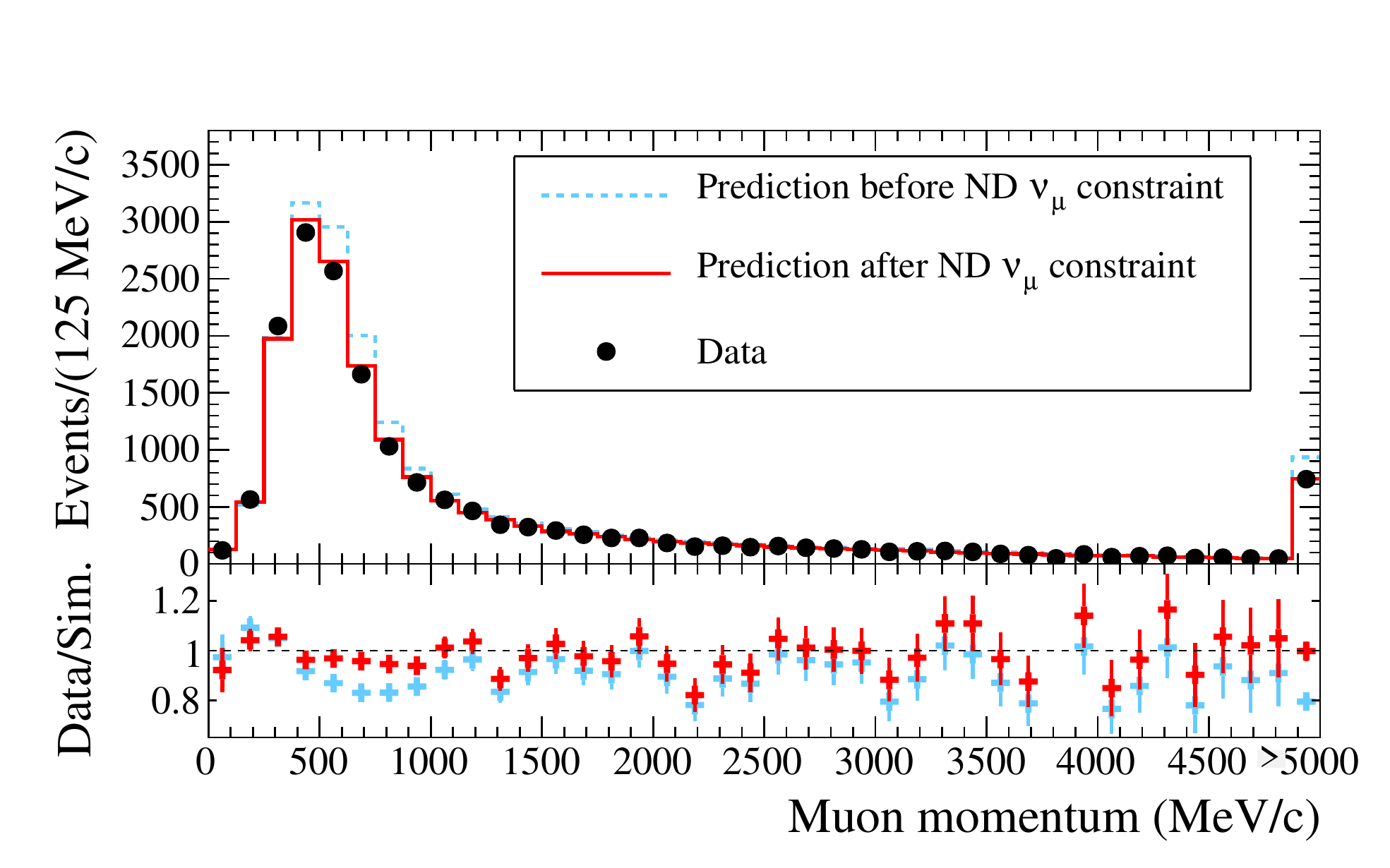}
\includegraphics[width=\linewidth]{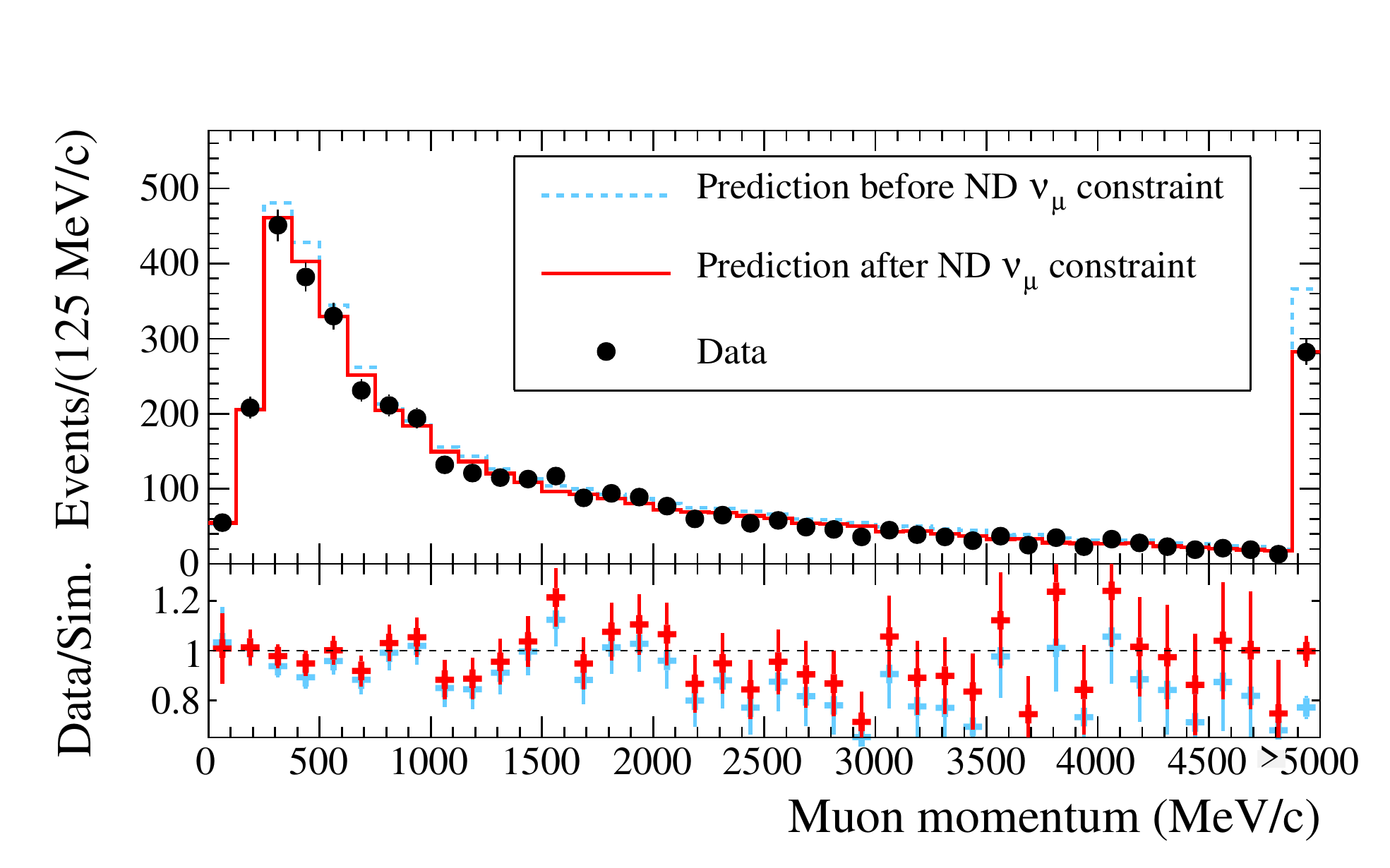}
\end{center}
\caption{Reconstructed momentum distribution for events selected in the ND280 $\nu_{\mu}$ CC analysis in the CC0$\pi$ sample (top) and in the CC1$\pi$ sample (bottom) before and after the fit to evaluate flux and cross section parameters. The last bins contain all the events with reconstructed muon momentum larger than 5~\gev/c. The error on the points is the statistical error on the data.}
\label{fig:numubanff}
\end{figure}

The muon momentum distributions for the CC0$\pi$ and CC1$\pi$ samples are shown in Fig.~\ref{fig:numubanff} together with the prediction before and after the fit.

The systematic parameters and their uncertainties obtained from the fit to the ND280 \num CC data are shown in Tab.~\ref{tab:banffpar}. The effect of those systematic uncertainties on the evaluation of the beam $\nu_e$ component is 6\%, as will be shown in Sect.~\ref{sec:results}. Without using the ND280 \num data the uncertainty on the beam $\nu_e$ prediction is larger than 20\%.

In the analysis described in this paper, systematic uncertainties of final state interactions are not evaluated from the ND280 $\nu_{\mu}$ CC samples; their effect is evaluated in bins of reconstructed electron momentum using the pion scattering data described in Sect.~\ref{sec:xsec}. 

\subsection{\label{sec:systsummary} Summary of the systematic uncertainties}
 The 60 parameters included in this analysis are shown in Tab.~\ref{tab:banffpar}. The variance due to limited statistics in the simulation is
added to the diagonal of the detector and FSI parameter covariance matrix.\\

\begin{table}[htbp]
\begin{ruledtabular}
\caption{\label{tab:banffpar} Summary of the values and errors of the systematic parameters used as priors for the measurement described in this paper. The values in squared brackets show the range of variation for systematic uncertainties with more than one parameter as a function of neutrino energy.}
\begin{tabular}{cccc}
Systematic & Central & Uncertainty & Number  \\
& Value & & of parameters \\
\hline
$\nu_{\mu}$-flux & [0.93-1.05] & [0.07-0.08] & 11 \\
$\nu_{e}$-flux & [0.95-1.02] & [0.07-0.09] & 7 \\
$\bar{\nu}_{\mu}$-flux & [0.99-1.03] & [0.09-0.14] & 5 \\
$\bar{\nu}_{e}$-flux & [0.95-1.01] & [0.08-0.17] &  2 \\
$M_{A}^{QE} [\unit{GeV}] $ & 1.24 & 0.07 & 1\\
$x_{1,2,3}^{QE}$ & [0.85-0.97] & [0.07-0.11] & 3  \\
$x_{SF}$ & 0.24 & 0.13 & 1 \\
$p_F[\unit{MeV/c}]$ & 266 & 11 & 1 \\
$E_b[\unit{MeV}]$ & 30.9 & 5.2 & 1 \\
$M_{A}^{RES} [\unit{GeV}] $ & 0.96 & 0.07 & 1\\
$x_{1,2}^{\mathrm{CC1}\pi}$ & [1.12-1.26] & [0.16-0.17] & 2 \\
$x^{\mathrm{NC1}\pi^0}$ &  1.14 & 0.25 & 1   \\
$x^{\mathrm{NC~other}}$ & 1.41 & 0.22 & 1 \\
$x^{\pi\mathrm{-less}}$ & 0.21 & 0.09 & 1 \\
$x^{\mathrm{CC~coh.}}$  & 0.45 & 0.16 & 1 \\
$x^{\mathrm{CC~other}}$ (GeV) & 0.23 & 0.29 & 1 \\
$\sigma_{\bar{\nu}}/\sigma_{\nu}$ & 1 & 0.4 & 1 \\
Detector+FSI & 1 & [0.07-0.19] & 17 \\
Out FGD FV $e^{-}$	 & $1$ & $0.3$  & 1 \\
Out FGD FV oth.	 & $1$ & $0.3$  & 1 \\
\hline
\boldmath{Total} & & & 60 \\
\end{tabular} 
\end{ruledtabular}

\end{table} 	

\section{\label{sec:fit} Measurement of the electron neutrino
  component of the beam}
In order to extract the beam $\nu_e$ component, a likelihood
fit of the selected distributions (CCQE-like, CCnonQE-like and
$\gamma$ samples) has been performed. To account for systematic
uncertainties, 60 nuisance parameters are included in the likelihood
function.
An additional term, $R(\nu_e)$, is included that scales the expected
number of events arising from \nue interactions. A second approach
introduces two additional terms, $R(\nue(\mu))$ and $R(\nue(K))$, which separately
scale the expected number of \nue events arising from muon and kaon
decays respectively.

It is important to notice that in this analysis the cross section parameters shown in
Tab.~\ref{tab:banffpar} have been evaluated from the ND280
$\nu_{\mu}$ CC samples assuming that the effective cross section
parameters for \num and \nue interactions are the same. A
measurement of the \nue CC interactions at ND280 is important to
put experimental constraints on this assumption that is poorly
constrained from experimental data~\cite{Day:2012gb}.

The likelihood ratio applied to the momentum distributions of the
$\nu_e$ CCQE-like, $\nu_e$ CCnonQE-like and $\gamma$ samples is:
\begin{align}
-2\ln \lambda (R;\vec{f}) = & - 2\ln \lambda_{\nu_eCCQE}(R;\vec{f}) \notag\\
                                            & - 2\ln \lambda_{\nu_e CCnonQE}(R;\vec{f}) \notag\\
                                            & - 2\ln \lambda_{\gamma}(R;\vec{f}) \notag\\
                                            & + (\vec{f} - \vec{f}_0)^T V ^{-1} (\vec{f} - \vec{f}_0) \notag\\
\end{align}
where $- 2\ln \lambda_{i}$ is the likelihood ratio
for each sample and the last term is the penalty term that
constrains the systematic nuisance parameters.
$\vec{f}_0$ is the vector of central values for the systematic
parameters shown in Tab.~\ref{tab:banffpar}, $\vec{f}$ is the vector of
nuisance parameters and $V$ is the covariance matrix that takes into account
the correlations among the systematics parameters.

The likelihood function is calculated in 18 bins of reconstructed electron
momentum from 200 MeV/c to 10 GeV/c for each selection, with a 100
MeV/c bin width up to 1 GeV/c and a wider binning at higher momenta.

The predicted number of events in each bin depends on the free
parameter $R(\nu_e)$ and on the 60 systematic
parameters according to the formula:

\begin{align}
n_{exp}^i (R;\vec{f}) = \frac{N^{data}_{\pot}}{N^{sim.}_{\pot}} \cdot \sum_j^{N_i} &
f^j_{flux} f^j_{x(norm.)} \notag\\
                                                                            & f^j_{x(resp.)} f^j_{det}(p_i) \times R(\nu_e)
\label{eq:nexp}
\end{align}
The sum $j$ runs over the events with momentum compatible
with the $i$-th momentum bin $p_i$, with $N_i$ total events in that bin. $N^{data}_{\pot}$ and $N^{sim.}_{\pot}$ are the
number of \pot for data and simulation. $f_{flux}$ multiply the flux
prediction in bins of true neutrino energy. $f_{det}$ multiply the
expected number of events in bins of electron reconstructed momentum. The cross section parameters are treated in
two ways: $f_{x(norm.)}$ multiply
the cross section normalization for a given true neutrino
energy and interaction model while $f_{x(resp.)}$ are pre-calculated response functions whose
value for the nominal parameter settings of Tab.~\ref{tab:banffpar}
is one. The reason for which they are treated in this way is that they can have a
non-linear dependency on the cross section parameters.
The prediction for the number of 
selected events before and after the application of the detector
systematic and of the flux and cross section
parameters obtained from the fit to the ND280 $\nu_{\mu}$ CC samples
is shown in Tab.~\ref{tab:nevents}.

For each of the three samples $k$ the likelihood ratio is:
\begin{align}
-2\ln \lambda_{k} (R;\vec{f}) =  2\times & \sum _{i=1}^{18} \left\lbrace \right.  n^i_{exp}(R;\vec{f}) - n^i_{data} \notag\\
                                 & \left. + n^i_{data} \ln \left(n^i_{data}/n^i_{exp}(R;\vec{f})\right) \right\rbrace \notag\\
\end{align}
where $n^i_{exp}$ and $n^i_{data}$ are the number of
expected and observed events in the $i$-th momentum bin.

\begin{table}
\caption{\label{tab:nevents} Number of events for CCQE-like and
  CCnonQE-like selections before and after the fit to the ND280 \num CC samples.}
\begin{ruledtabular}
 \begin{tabular}{ldddddd}
 \multirow{2}{*}{Category} & \multicolumn{2}{c}{CCQE selection} & \multicolumn{2}{c}{CCnonQE selection} \\
                           & \multicolumn{1}{c}{Before} &
                           \multicolumn{1}{c}{After}
                           & \multicolumn{1}{c}{Before} & \multicolumn{1}{c}{After}  \\
\hline 
 \nue from $\mu$          & 61.2 & 52.7       & 38.9 & 34.9\\
 \nue from K              &125.9 & 108.2      &253.6 & 221.0\\
 $\nu_{\mu}$ (in FV) & 30.4 & 31.7       &124.9 & 132.4\\
 $\nu_{\mu}$ (OOFV)  & 58.1 & 67.0       & 27.5 & 26.8\\
\hline
Total & 275.5 & 259.6 & 444.9 & 415.1 \\
\hline
\bf{Data} & \multicolumn{2}{d}{225} &  \multicolumn{2}{d}{392} \\ 
 \end{tabular}
\end{ruledtabular}
\end{table}

\section{\label{sec:results} Results}
The result obtained for the parameter $R(\nu_e)$ is:

\begin{align}
R(\nu_e) = 1.01 & \pm 0.06 (stat.) \pm 0.06 (flux \oplus x.sec)
\notag \\
  & \pm 0.05 (det. \oplus FSI) = 1.01 \pm 0.10
\end{align}

where the first term represents the statistical error, the second term
represents the systematic uncertainties related to the flux and to
the cross section models, and the last term represents the
detector systematics and the FSI uncertainties.

This result indicates that the beam $\nu_e$ component measured in the
data is consistent with the expectation for this component after
the constraint from the ND280 \num CC sample. This is a key validation of the
strategy followed by T2K to constrain the flux and cross section parameters
for all the neutrino oscillation analyses.

The second measurement is performed by fitting independently the
$\nu_e$ originating from muons and from kaons. As is shown in
Fig.~\ref{fig:t2kflux} the $\nu_e$ from muons mainly populate the low
energy region while the $\nu_e$ from kaons are dominant at high
momenta. 

Given the larger efficiency of the analysis at high energy (shown in
Fig.~\ref{fig:nue_eff_inc_e}) there are three times as many selected
events from kaon decay as from muon decay. The \nue from muon decay mainly populate the
CCQE-like selection in its low momentum region.
The results obtained for the two components are:

\begin{align}
R(\nu_e(\mu)) = 0.68 & \pm 0.24 (stat.) \pm 0.11 (flux \oplus x.sec)
\notag \\
& \pm 0.14 (det. \oplus FSI) = 0.68 \pm 0.30
\end{align}

and

\begin{align}
R(\nu_e(K)) = 1.10 & \pm 0.08  (stat.) \pm 0.09 (flux \oplus x.sec)
\notag \\
 & \pm 0.06 (det. \oplus FSI) = 1.10 \pm 0.14 
\end{align}

\begin{figure} [!h]
\begin{center}
\includegraphics[width=\linewidth]{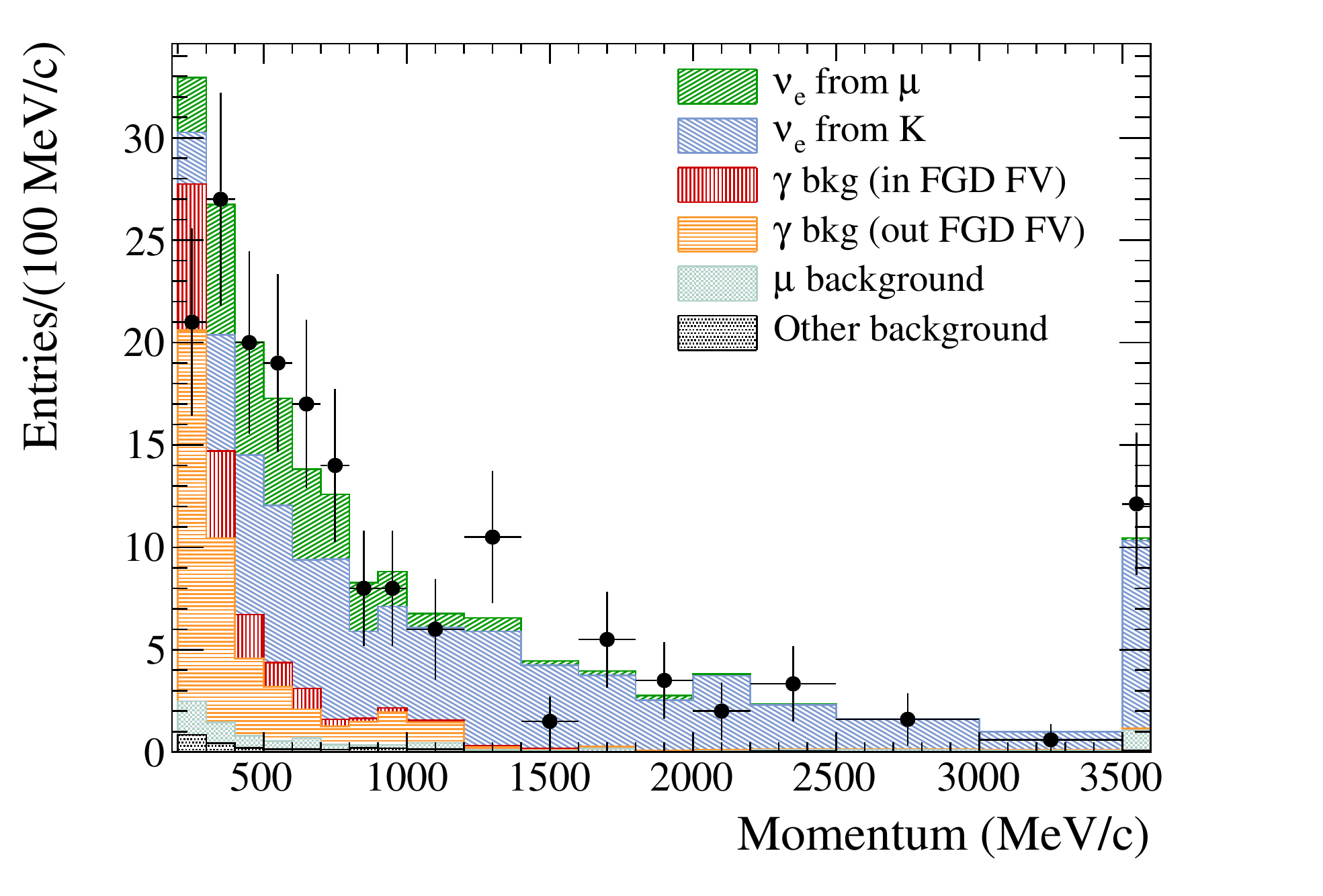}
\includegraphics[width=\linewidth]{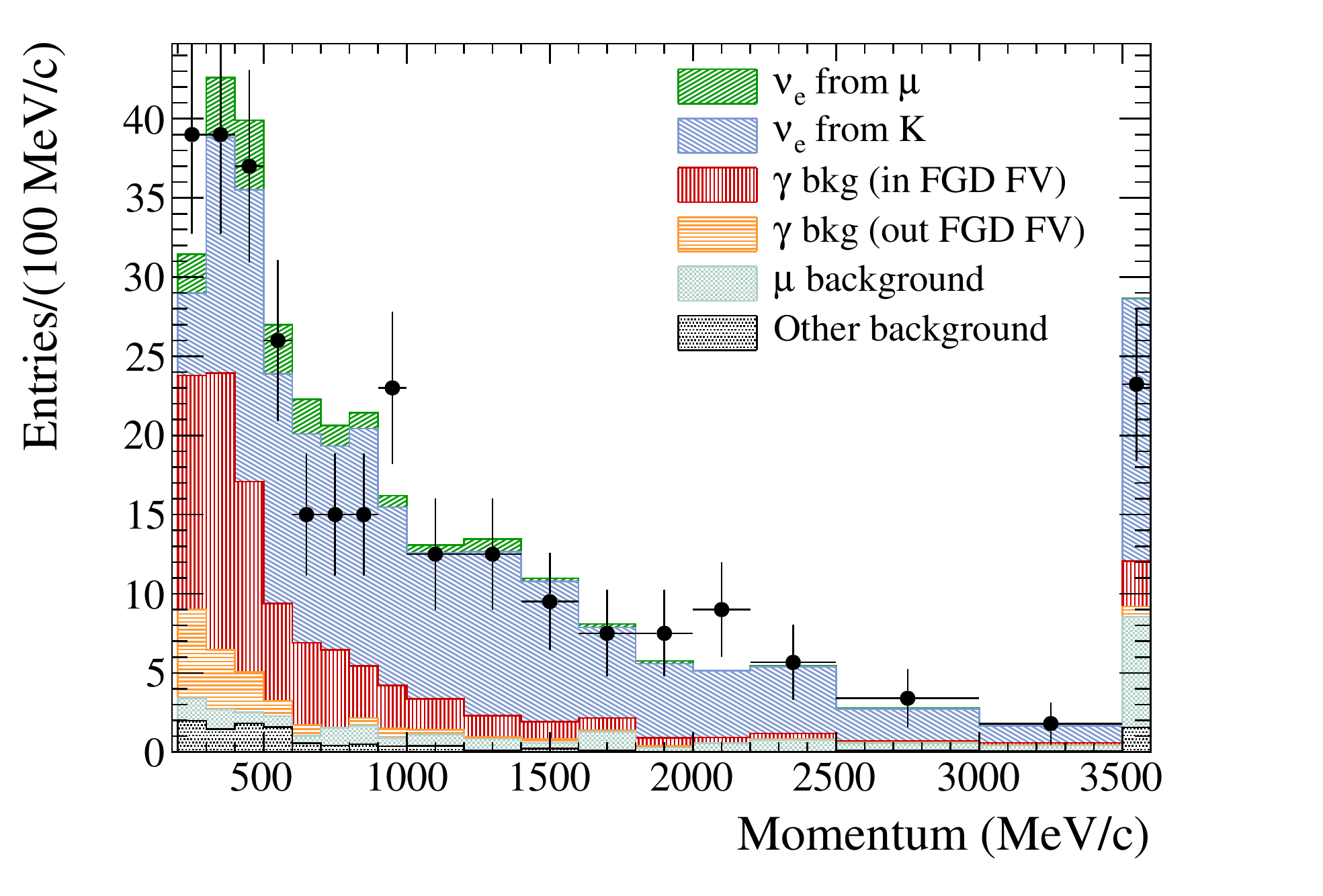}
\includegraphics[width=\linewidth]{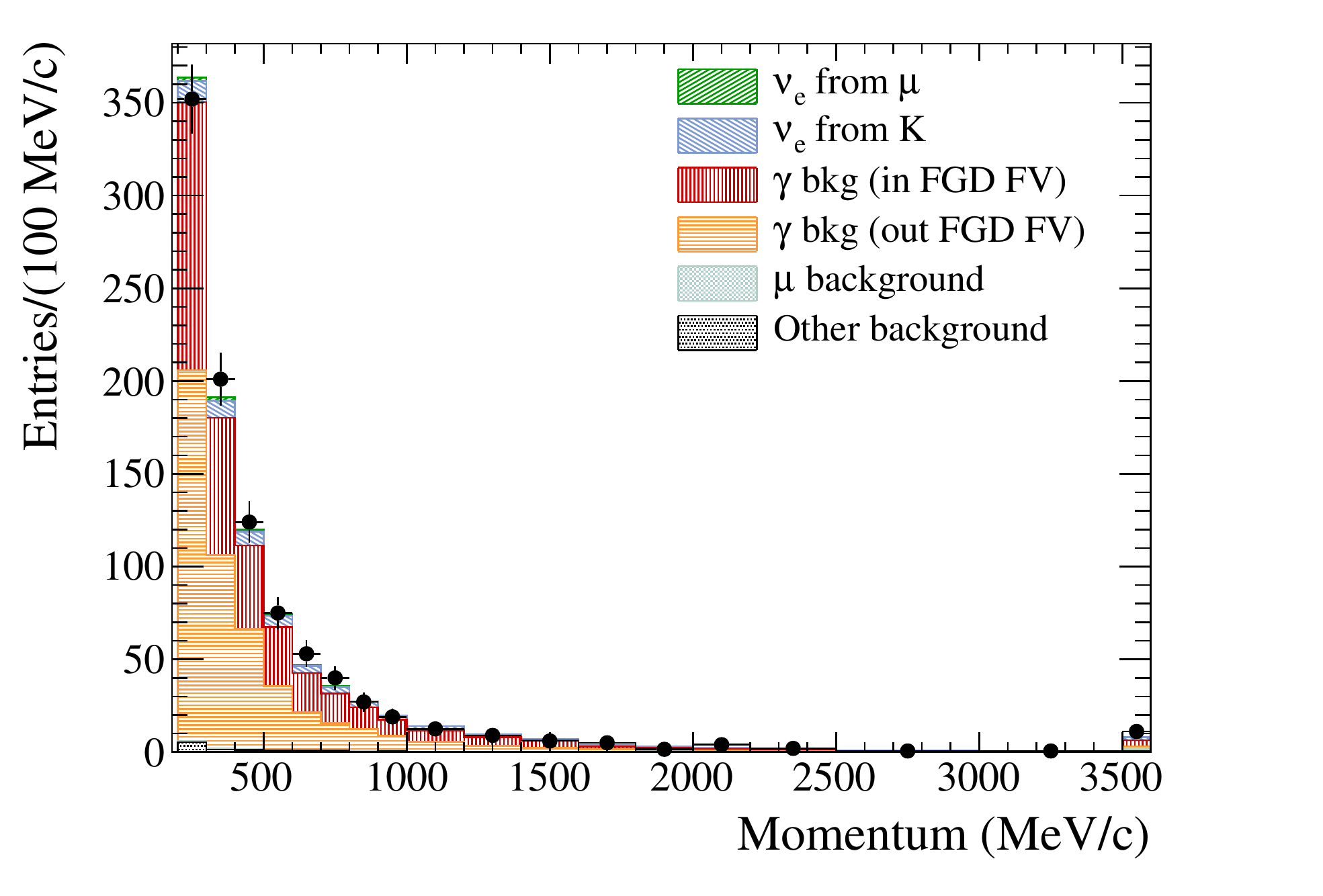}
\end{center}
\caption{Reconstructed electron momentum distribution for the events selected
  in the three samples after the fit to extract the beam
  \nue component: CCQE-like selection (top),
  CCnonQE-like selection (center) and $\gamma$ selection (bottom). The
last bin contains all the events with reconstructed electron momentum
larger than 3.5 \gev/c. The signal is divided into \nue produced by
muon and kaon decays. The background is divided into the same categories as Fig.~\ref{fig:nue_postpid_reac}. The error on the points is the statistical error on the data.}
\label{fig:results}
\end{figure}

Due to the small amount of \nue coming from muons,
the uncertainty on the measurement of this component is still
statistically limited and will be improved when more data is
available. With the present statistics both numbers are compatible
with unity showing no discrepancies between the predicted and the
observed beam \nue component.
The larger systematic uncertainty for $R(\nu_e(\mu))$ is due
to the fact that the detector, flux and cross-section systematic
uncertainties are larger at low momenta.
The distribution of the reconstructed electron momentum for the three samples after the fit are shown in
Fig.~\ref{fig:results}. 

As far as the nuisance parameters are concerned, the fitted values are
in good agreement with the expectations. The out
of FGD electron component is reduced in the fit by $0.64\pm0.10$, compatible with the prior systematic
uncertainty of 30\%.
This reduction might point to the fact that the simulation does
not properly reproduce the amount of $\pi^0$ produced in neutrino
interactions in the materials surrounding the ND280 tracker
region. Those interactions are mainly high energy deep inelastic
scattering events for which the $\pi^0$ multiplicity is not well measured. This reduction does not have a large impact on the
measurements presented here because of the presence of the photon
conversion sample used to evaluate this background.

\section{\label{sec:summary} Summary}

In summary, a selection of \nue CC interactions has been performed using
the T2K off-axis near detector combining the PID capabilities
of the TPC and ECal.  The combination of these two detectors allows the selection of a clean sample
of electrons with a purity of 92\% and a muon misidentification
probability smaller than 1\%.

The selected sample is mainly composed of electrons coming from
\nue CC interactions but a non negligible component comes
from photon conversions in the FGD. This background is constrained
in the analysis using a sample of \epem pairs coming from photon conversions in which both
outgoing particles are reconstructed in the TPC.

To extract the beam \nue component from the data a likelihood fit is performed. The expected number of \nue interactions 
is predicted by the same model used for the T2K oscillation
analyses where the neutrino fluxes and the neutrino cross sections are 
evaluated by the \num CC samples selected at ND280.

The observed number of events is in good agreement with the
prediction, providing a direct confirmation of this method. This
measurement is still statistically limited but when additional data
is collected it will be possible to further improve the
measurement of the intrinsic \nue component in the T2K beam and perform
measurements of \nue cross sections and of the \nue/\num
cross section differences that have not been measured at T2K energies.

This measurement is particularly important because the intrinsic \nue component is the main
background for all the proposed long-baseline neutrino oscillation
experiments 
aiming to measure CP violation in the leptonic sector. In this paper it
is shown that, although the component is small, it is possible
to measure it with a properly designed near detector.

\begin{acknowledgments}
We thank the J-PARC staff for superb accelerator performance and the
CERN NA61 collaboration for providing valuable particle production
data. We acknowledge the support of MEXT, Japan; NSERC, NRC and CFI,
Canada; CEA and CNRS/IN2P3, France; DFG, Germany; INFN, Italy;
Ministry of Science and Higher Education, Poland; RAS, RFBR and MES,
Russia; MICINN and CPAN, Spain; SNSF and SER, Switzerland; STFC, U.K.;
and DOE, U.S.A. We also thank CERN for the UA1/NOMAD magnet, DESY for
the HERA-B magnet mover system, NII for SINET4, the WestGrid and
SciNet consortia in Compute Canada, and GridPP, UK. In addition
participation of individual researchers and institutions has been
further supported by funds from: ERC (FP7), EU; JSPS, Japan; Royal
Society, UK; DOE Early Career program, U.S.A.
\end{acknowledgments}

\bibliographystyle{unsrt}
\bibliography{beam_nue}


\end{document}